\documentclass[english,preprint,nofootinbib]{revtex4}
\usepackage[T1]{fontenc}
\usepackage[latin9]{inputenc}
\usepackage{color}
\usepackage{babel}
\usepackage{float}
\usepackage{amsmath}
\usepackage{amssymb}
\usepackage{graphicx}
\usepackage{esint}
\usepackage[unicode=true,bookmarks=false,breaklinks=false,pdfborder={0 0 1},backref=section,colorlinks=true]
 {hyperref}
\hypersetup{
 urlcolor=black,linkcolor=black,citecolor=black}
\usepackage{breakurl}

\makeatletter
\@ifundefined{textcolor}{}
{%
 \definecolor{BLACK}{gray}{0}
 \definecolor{WHITE}{gray}{1}
 \definecolor{RED}{rgb}{1,0,0}
 \definecolor{GREEN}{rgb}{0,1,0}
 \definecolor{BLUE}{rgb}{0,0,1}
 \definecolor{CYAN}{cmyk}{1,0,0,0}
 \definecolor{MAGENTA}{cmyk}{0,1,0,0}
 \definecolor{YELLOW}{cmyk}{0,0,1,0}
}

\usepackage{amsfonts}
\usepackage{babel}
\usepackage{graphicx}
\providecommand{\U}[1]{\protect\rule{.1in}{.1in}}
\@ifundefined{textcolor}{}{
\definecolor{BLACK}{gray}{0}
 \definecolor{WHITE}{gray}{1}
 \definecolor{RED}{rgb}{1,0,0}
 \definecolor{GREEN}{rgb}{0,1,0}
 \definecolor{BLUE}{rgb}{0,0,1}
 \definecolor{CYAN}{cmyk}{1,0,0,0}
 \definecolor{MAGENTA}{cmyk}{0,1,0,0}
 \definecolor{YELLOW}{cmyk}{0,0,1,0}
 }

\makeatother

\begin{document}

\title{Higher spin extension of cosmological spacetimes in 3D:\\
asymptotically flat behaviour with chemical potentials\\
and thermodynamics}

\author{Javier Matulich$^{1}$, Alfredo Pérez$^{1}$, David Tempo$^{1,2}$,
Ricardo Troncoso$^{1}$}

\email{matulich, aperez, tempo, troncoso@cecs.cl}

\affiliation{$^{1}$Centro de Estudios Científicos (CECs), Av. Arturo Prat 514,
Valdivia, Chile}

\affiliation{$^{2}$Physique Théorique et Mathématique Université Libre de Bruxelles
and International Solvay Institutes Campus Plaine C.P. 231, B-1050
Bruxelles, Belgium.}

\preprint{CECS-PHY-14/03}
\begin{abstract}
A generalized set of asymptotic conditions for higher spin gravity
without cosmological constant in three spacetime dimensions is constructed.
They include the most general temporal components of the gauge fields
that manifestly preserve the original asymptotic higher spin extension
of the BMS$_{3}$ algebra, with the same central charge. By virtue
of a suitable permissible gauge choice, it is shown that this set
can be directly recovered as a limit of the boundary conditions that
have been recently constructed in the case of negative cosmological
constant, whose asymptotic symmetries are spanned by two copies of
the centrally-extended W$_{3}$ algebra. Since the generalized asymptotic
conditions allow to incorporate chemical potentials conjugated to
the higher spin charges, a higher spin extension of locally flat cosmological
spacetimes becomes naturally included within the set. It is shown
that their thermodynamic properties can be successfully obtained exclusively
in terms of gauge fields and the topology of the Euclidean manifold,
which is shown to be the one of a solid torus, but with reversed orientation
as compared with one of the black holes. It is also worth highlighting
that regularity of the fields can be ensured through a procedure that
does not require an explicit matrix representation of the entire gauge
group. In few words, we show that the temporal components of generalized
dreibeins can be consistently gauged away, which partially fixes the
chemical potentials, so that the remaining conditions can just be
obtained by requiring the holonomy of the generalized spin connection
along a thermal circle to be trivial. The extension of the generalized
asymptotically flat behaviour to the case of spins $s\geq2$ is also
discussed.
\end{abstract}
\maketitle

\section{Introduction}

Higher spin gravity in three-dimensional spacetimes has become the
source of a great deal of activity; see e.g., \cite{PTT-GGS, PTT-H1, PTT-CFPT2,PTT-PTT1,PTT-PTT2,PTT-CS,PTT-CJS,PTT-GHJ,PTT-F-GK1,PTT-F-GK2, PTT-F-GK3z, PTT-KU, PTT-dBJ, PTT-dBJ2, PTT-ACI, PTT-F-GK3, PTT-F-GK4, PTT-Last-1, PTT-Last, PTT-LLW, PTT-F-PTT1, PTT-G-Lif, PTT-FN, PTT-PK, PTT-TAN, PTT-GGR, PTT-BCT, PTT-FHK, PTT-Super-HS, PTT-Super-1, PTT-Super-SD, PTT-Super-T, PTT-CGGR, PTT-H4, PTT-H6, PTT-H7, PTT-CPR, PTT-GJP, PTT-CF, Afshar:2014cma, Beccaria:2014jxa, Riegler:2014bia, Grumiller:2014lna, Craps:2014wpa, Kiran:2014kca}.
Reviews about this subject can be found in refs. \cite{PTT-Rew-AGKP},
\cite{PTT-Rew-J}, \cite{ReviewHS4}, \cite{Campoleoni:2011tn}, \cite{PTT}.
In the case of negative cosmological constant, black hole solutions
endowed with global higher spin charges have been recently described
in \cite{HPTT}, \cite{BHPTT}. The theory can be naturally formulated
in terms of a Chern-Simons action \cite{PTT-3D-2}, \cite{PTT-3D-3},
\cite{PTT-3D-4}, so that in the simplest case, the gauge group is
given by two copies of $SL(3,%TCIMACRO{\U{211d}}%
%BeginExpansion
\mathbb{R}%EndExpansion
)$, and it describes nonpropagating spin-$3$ fields nonminimally coupled
to AdS$_{3}$ gravity. As shown in \cite{BHPTT}, the theory admits
two different classes of black hole solutions, due to the fact that
there are two inequivalent sets of asymptotic conditions whose asymptotic
symmetry algebra corresponds to different extensions of the conformal
group in two dimensions. In one case, the asymptotic symmetry algebra
is spanned by two copies of the centrally-extended W$_{3}^{(2)}$
algebra, so that the black holes within this sector, apart from the
mass and the angular momentum, can be endowed with $U(1)$ and bosonic
spin-$3/2$ charges. The black hole solutions previously found in
\cite{PTT-GK}, \cite{PTT-AGKP}, \cite{PTT-CM}, were shown to correspond
to particular cases thereof. Hereafter, we will focus in the remaining
case, in which the asymptotic symmetries are generated by two copies
of the W$_{3}$ algebra with the same central extension as in the
case of pure gravity on AdS \cite{Brown-Henneaux}. This sector includes
higher spin black holes that generalize the BTZ solution \cite{BTZ},
\cite{BHTZ} to include spin-$3$ charges. As explained in section
\ref{section_limit}, there is a suitable gauge choice that allows
to perform the vanishing cosmological constant limit in a straightforward
way, so that the higher spin black hole solution reduces to a higher
spin extension of locally flat cosmological spacetimes \cite{Ezawa:1992nk},\cite{Cornalba:2002fi},\cite{Cornalba:2003kd},\cite{Barnich:2012xq},\cite{Bagchi:2012xr}.
It is found that the latter class of solutions does not fit within
the set of asymptotic conditions describing asymptotically flat spacetimes
in the context of higher spin gravity, independently proposed in \cite{ABFGR},
\cite{GMPT}. Hence, one of the purposes of this article is to extend
these asymptotic conditions from scratch, so as to include the higher
spin extension of locally flat cosmological spacetimes, without spoiling
the original asymptotic symmetries, generated by a higher spin extension
of the BMS$_{3}$ algebra with central charge. This is explicitly
carried out along the lines of \cite{HPTT}, \cite{BHPTT}, in section
\ref{section-flat conditions}. We next show in section \ref{section_limit}
that the special gauge choice aforementioned, actually allows to recover
the whole asymptotic structure from the one proposed in \cite{BHPTT}
in the vanishing cosmological constant limit. Since the generalized
asymptotic conditions incorporate the chemical potentials conjugated
to the higher spin charges, the thermodynamic properties of the higher
spin extension of the cosmological spacetimes can be readily analyzed.
This is performed in section \ref{HScosmo+thermo}, where we start
warming up with the pure gravity case, and then we show how to switch
on the higher spin charges and their corresponding chemical potentials.
It is worth highlighting that the thermodynamic properties can be
suitably analyzed without the need of an explicit matrix representation
of the entire gauge group. Finally, section \ref{final remarks} is
devoted to final remarks, including the extension to fields of spin
$s\geq2$.

It must be pointed out that many our results overlap with the ones
recently found by Gary, Grumiller, Riegler and Rosseel \cite{Gary:2014ppa}.
Both approaches were carried out independently, and so they turn out
to be radically different. Nonetheless, as it is discussed in section
\ref{final remarks}, it is reassuring to verify that our results
agree in the cases that were considered in \cite{Gary:2014ppa}.

\section{Generalized asymptotically flat behaviour}

\label{section-flat conditions}

Higher spin gravity in three spacetime dimensions is remarkably simpler
than its higher-dimensional counterparts \cite{PTT-VV1}, \cite{PTT-VV2}.
Indeed, in this case the theory can be consistently truncated in order
to describe the dynamics of a finite number of fields with spin $s=2,3,...,N$
\cite{PTT-3D-2}, \cite{PTT-3D-3}, \cite{PTT-3D-4}. Furthermore, unlike
the case of higher dimensions, the three-dimensional theory also admits
a suitable formulation with vanishing cosmological constant. Afterwards,
for the sake of simplicity, we will focus in the case of $s=2$, $3$,
so that we leave the extension to fields of spins $s\geq2$ to be
depicted in section \ref{final remarks}.

The theory is described by a Chern-Simons action, given by 
\begin{equation}
I_{CS}[A]=\frac{k}{4\pi}\int\langle AdA+\frac{2}{3}A^{3}\rangle\ ,\label{CS2}
\end{equation}
where the gauge field $A=A_{\mu}dx^{\mu}$ reads (see e.g., \cite{Theisen-HS})
\begin{equation}
A=\omega^{a}J_{a}+e^{a}P_{a}+W^{ab}J_{ab}+E^{ab}P_{ab}\ ,\label{A=00003Dw+e}
\end{equation}
and the set $\left\{ J_{a},P_{a},J_{ab},P_{ab}\right\} $ that spans
the gauge group, is such that the generators $P_{ab}$, $J_{ab}$
are assumed to be symmetric and traceless. In the case of vanishing
cosmological constant, as explained in \cite{Theisen-HS}, the corresponding
Lie algebra turns out to be a generalization of the Poincaré algebra,
whose nonvanishing commutators read
\begin{align}
\lbrack J_{a},J_{b}] & =\epsilon_{abc}J^{c}\ \ ;\ \ [J_{a},P_{b}]=\epsilon_{abc}P^{c}\ \ ,\nonumber \\
\lbrack J_{a},J_{bc}] & =\epsilon_{\;\; a(b}^{m}J_{c)m}\ \ ;\ \ [J_{a},P_{bc}]=\epsilon_{\;\; a(b}^{m}P_{c)m}\ \ ;\ \ [P_{a},J_{bc}]=\epsilon_{\;\; a(b}^{m}P_{c)m}\ \ ,\label{HS_Poincare}\\
\lbrack J_{ab},J_{cd}] & =-\eta_{\left(a\right\vert (c}\epsilon_{d)\left\vert b\right)m}J^{m}\ \ ;\ \ [J_{ab},P_{cd}]=-\eta_{\left(a\right\vert (c}\epsilon_{d)\left\vert b\right)m}P^{m}\ \ ,\nonumber 
\end{align}
which is naturally recovered through an Inönü-Wigner contraction of
two copies of $sl(3,%TCIMACRO{\U{211d}}%
%BeginExpansion
\mathbb{R}%EndExpansion
)$. In (\ref{CS2}), the level is determined by the Newton constant
according to $k=\frac{1}{4G}$, and the non-degenerate invariant bilinear
product, is such that the only nonvanishing components of the bracket
are given by
\begin{equation}
\langle P_{a}J_{b}\rangle=\eta_{ab}\ \ \ ,\ \ \ \langle P_{ab}J_{cd}\rangle=\eta_{ac}\eta_{bd}+\eta_{ad}\eta_{cb}-\frac{2}{3}\eta_{ab}\eta_{cd}\ .\label{CS3}
\end{equation}
The field equations are then solved by locally flat connections, fulfilling
$F=dA+A^{2}=0$.

A consistent set of boundary conditions for this theory was proposed
independently in \cite{ABFGR}, \cite{GMPT}, whose asymptotic symmetries
were shown to be spanned by a higher spin extension of the BMS$_{3}$
algebra with an appropriate central extension. In \cite{GMPT}, it
was also shown that there is a suitable gauge choice that allows to
successfully recover the whole asymptotic structure from the one proposed
independently in \cite{Theisen-HS}, \cite{Henneaux-HS} in the vanishing
cosmological constant limit.

In refs. \cite{ABFGR}, \cite{GMPT} it was also argued, along different
lines, that it would be worth exploring whether the asymptotic conditions
could be extended in a consistent way with the asymptotic BMS$_{3}$
symmetry. Indeed, soon after it was shown in \cite{HPTT}, \cite{BHPTT}
that this task can always be achieved in a systematic way and for
a generic setting, where some explicit examples were constructed in
the case of negative cosmological constant.

In what follows, we explain how the asymptotic conditions presented
in \cite{ABFGR}, \cite{GMPT} can be generalized along the lines
of \cite{HPTT}, \cite{BHPTT}, so as to include chemical potentials
without spoiling the original asymptotic BMS$_{3}$ symmetry.

Let us then start considering the asymptotic form of the gauge fields
in \cite{GMPT} at a fixed time slice ($u=u_{0}$). It is useful to
express it with the gauge choice of \cite{BDMT}, which allows to
completely capture the radial dependence through a group element of
the form $h(r)=e^{-rP_{0}}$, so that
\begin{equation}
A=h^{-1}a_{(0)}h+h^{-1}dh\ ,\label{A-Flat}
\end{equation}
where $a_{(0)}=a_{\varphi}d\varphi$, with%
\footnote{Our conventions agree with the ones in \cite{GMPT}, being such that
a non-diagonal Minkowski tangent space metric is assumed, whose only
nonvanishing components are $\eta_{01}=\eta_{10}=\eta_{22}=1$, and
the Levi-Civita symbol fulfils $\epsilon_{012}=1$. Nontheless, the
fields used in \cite{GMPT}, here denoted with tilde, relate with
the ones in this work according to $\mathcal{P}=\frac{k}{4\pi}\mathcal{\tilde{M}}$,
$\mathcal{J}=\frac{k}{2\pi}\mathcal{\tilde{N}}$, $\mathcal{W}=\frac{k}{\pi}\mathcal{\tilde{W}}$,
$\mathcal{V}=\frac{k}{\pi}\mathcal{\tilde{Q}}$. For later purposes,
it is useful to introduce the subscript $(0)$ in order to denote
objects defined in the case of vanishing cosmological constant $\Lambda$,
e.g., $a_{(0)}=a_{(\Lambda=0)}$.%
}
\begin{equation}
a_{\varphi}=J_{1}+\frac{2\pi}{k}(\mathcal{J}P_{0}+\mathcal{P}J_{0})+\frac{\pi}{k}(\mathcal{V}P_{00}+\mathcal{W}J_{00})\ .\label{a-phi-flat}
\end{equation}
The asymptotic form of the connection is then maintained under gauge
transformations of the form $\delta A=d\Omega+\left[A,\Omega\right]$,
with $\Omega=h^{-1}\eta_{(0)}h$, where $\eta_{(0)}=\eta_{(0)}\left(T,Y,Z,X\right)$
depends on four arbitrary functions of $u_{0}$, $\varphi$, and reads
\begin{align}
\eta_{(0)} & =\frac{2\pi}{k}\left(Y\mathcal{J}+T\mathcal{P}+2Z\mathcal{V}+2Z\mathcal{W}-\frac{k}{2\pi}T^{\prime\prime}\right)P_{0}+TP_{1}-T^{\prime}P_{2}\nonumber \\
 & +\frac{2\pi}{k}\left(Y\mathcal{P}+2Z\mathcal{W}-\frac{k}{2\pi}\mu^{\prime\prime}\right)J_{0}+YJ_{1}-Y^{\prime}J_{2}+\frac{\pi}{k}\left[Y\mathcal{V}+T\mathcal{W}-\frac{8}{3}\left(X^{\prime\prime}\mathcal{P}+Z^{\prime\prime}\mathcal{J}\right)\right.\nonumber \\
 & \left.-\frac{7}{3}\left(Z^{\prime}\mathcal{J}^{\prime}+X^{\prime}\mathcal{P}^{\prime}\right)+\frac{2}{3}Z\left(\frac{12}{k}\mathcal{J}\mathcal{P}-\mathcal{J}^{\prime\prime}\right)+\frac{2}{3}X\left(\frac{6\pi}{k}\mathcal{P}^{2}-\mathcal{P}^{\prime\prime}\right)+\frac{k}{6\pi}X^{\prime\prime\prime\prime}\right]P_{00}\label{lambda tilda}\\
 & +\frac{4\pi}{k}\left(X\mathcal{P}+Z\mathcal{J}-\frac{k}{4\pi}X^{\prime\prime}\right)P_{01}-\frac{4\pi}{3k}\left(Z\mathcal{J}^{\prime}+X\mathcal{P}^{\prime}+\frac{5}{2}\left(X^{\prime}\mathcal{P}+Z^{\prime}\mathcal{J}\right)-\frac{k}{4\pi}X^{\prime\prime\prime}\right)P_{02}\nonumber \\
 & +XP_{11}-X^{\prime}P_{12}+\frac{\pi}{k}\left[Y\mathcal{W}+\frac{2}{3}Z\left(\frac{6\pi}{k}\mathcal{P}^{2}-\mathcal{P}^{\prime\prime}\right)-\frac{8}{3}Z^{\prime\prime}\mathcal{P}-\frac{7}{3}Z^{\prime}\mathcal{P}^{\prime}+\frac{k}{6\pi}Z^{\prime\prime\prime\prime}\right]J_{00}\nonumber \\
 & +\left(\frac{4\pi}{k}Z\mathcal{P}-Z^{\prime\prime}\right)J_{01}-\frac{4\pi}{3k}\left(Z\mathcal{P}^{\prime}+\frac{5}{2}Z^{\prime}\mathcal{P}-\frac{k}{4\pi}Z^{\prime\prime\prime}\right)J_{02}+ZJ_{11}-Z^{\prime}J_{12}\,,\nonumber 
\end{align}
provided the fields $\mathcal{P}$, $\mathcal{J}$, $\mathcal{W}$
and $\mathcal{V}$ transform according to
\begin{align}
\delta_{(0)}\mathcal{J} & =2Y^{\prime}\mathcal{J}+Y\mathcal{J}^{\prime}+T\mathcal{P}^{\prime}+2T^{\prime}\mathcal{P}-\frac{k}{2\pi}T^{\prime\prime\prime}+2Z\mathcal{V}^{\prime}+3Z^{\prime}\mathcal{V}+2\mathcal{W}^{\prime}X+3\mathcal{W}X^{\prime}\,,\nonumber \\
\delta_{(0)}\mathcal{P} & =2Y^{\prime}\mathcal{P}+Y\mathcal{P}^{\prime}-\frac{k}{2\pi}Y^{\prime\prime\prime}+2Z\mathcal{W}^{\prime}+3Z^{\prime}\mathcal{W}\,,\label{delta0}\\
\delta_{(0)}\mathcal{W} & =3Y^{\prime}\mathcal{W}+Y\mathcal{W}^{\prime}-\frac{2}{3}Z\left(\mathcal{P}^{\prime\prime}-\frac{8\pi}{k}\mathcal{P}^{2}\right)^{\prime}-3Z^{\prime}\left(\mathcal{P}^{\prime\prime}-\frac{32\pi}{9k}\mathcal{P}^{2}\right)\nonumber \\
 & -5Z^{\prime\prime}\mathcal{P}^{\prime}-\frac{10}{3}Z^{\prime\prime\prime}\mathcal{P}+\frac{k}{6\pi}Z^{\left(5\right)}\,,\nonumber \\
\delta_{(0)}\mathcal{V} & =3Y^{\prime}\mathcal{V}+Y\mathcal{V}^{\prime}+T\mathcal{W}^{\prime}+3T^{\prime}\mathcal{W}-\frac{2}{3}Z\left(\mathcal{J}^{\prime\prime}-\frac{16\pi}{k}\mathcal{J}\mathcal{P}\right)^{\prime}\nonumber \\
 & -3Z^{\prime}\left(\mathcal{J}^{\prime\prime}-\frac{64\pi}{9k}\mathcal{J}\mathcal{P}\right)-5Z^{\prime\prime}\mathcal{J}^{\prime}-\frac{10}{3}Z^{\prime\prime\prime}\mathcal{J}-\frac{2}{3}X\left(\mathcal{P}^{\prime\prime}-\frac{8\pi}{k}\mathcal{P}^{2}\right)^{\prime}\nonumber \\
 & -3X^{\prime}\left(\mathcal{P}^{\prime\prime}-\frac{32\pi}{9k}\mathcal{P}^{2}\right)-5\mathcal{P}^{\prime}X^{\prime\prime}-\frac{10}{3}\mathcal{P}X^{\prime\prime\prime}+\frac{k}{6\pi}X^{(5)}\,.\nonumber 
\end{align}
Here prime denotes derivative with respect to $\varphi$. Therefore,
as explained in \cite{HPTT}, \cite{BHPTT}, since the time evolution
of the dynamical fields is generated by a gauge transformation with
parameter $A_{u}$, the asymptotic symmetries will be preserved along
time provided the Lagrange multiplier is of the allowed form; i.e.,
$A_{u}=h^{-1}a_{u}h$, with
\begin{equation}
a_{u}=\eta_{(0)}\left(\xi,\mu,\vartheta,\varrho\right)\,,\label{au}
\end{equation}
where the ``chemical potentials\textquotedblright{} $\xi$, $\mu$,
$\vartheta$, $\varrho$, stand for arbitrary functions of time and
the angle that are fixed at the boundary.

Consistency of preserving the asymptotic form of $a_{u}$ under the
allowed gauge transformations then implies that the field equations
have to be fulfilled at the asymptotic region, i.e.,
\begin{align}
\dot{\mathcal{J}} & =2\mu^{\prime}\mathcal{J}+\mu\mathcal{J}^{\prime}+\xi\mathcal{P}^{\prime}+2\xi^{\prime}\mathcal{P}-\frac{k}{2\pi}\xi^{\prime\prime\prime}+2\vartheta\mathcal{V}^{\prime}+3\vartheta^{\prime}\mathcal{V}+2\mathcal{W}^{\prime}\varrho+3\mathcal{W}\varrho^{\prime}\ ,\nonumber \\
\dot{\mathcal{P}} & =2\mu^{\prime}\mathcal{P}+\mu\mathcal{P}^{\prime}-\frac{k}{2\pi}\epsilon^{\prime\prime\prime}+2\vartheta\mathcal{W}^{\prime}+3\vartheta^{\prime}\mathcal{W}\ ,\nonumber \\
\mathcal{\dot{W}} & =3\mu^{\prime}\mathcal{W}+\mu\mathcal{W}^{\prime}-\frac{2}{3}\vartheta\left(\mathcal{P}^{\prime\prime}-\frac{8\pi}{k}\mathcal{P}^{2}\right)^{\prime}-3\vartheta^{\prime}\left(\mathcal{P}^{\prime\prime}-\frac{32\pi}{9k}\mathcal{P}^{2}\right)\nonumber \\
 & -5\vartheta^{\prime\prime}\mathcal{P}^{\prime}-\frac{10}{3}\vartheta^{\prime\prime\prime}\mathcal{P}+\frac{k}{6\pi}\vartheta^{\left(5\right)}\label{feqs_flat}\\
\dot{\mathcal{V}} & =3\mu^{\prime}\mathcal{V}+\mu\mathcal{V}^{\prime}+\xi\mathcal{W}^{\prime}+3\xi^{\prime}\mathcal{W}-\frac{2}{3}\vartheta\left(\mathcal{J}^{\prime\prime}-\frac{16\pi}{k}\mathcal{J}\mathcal{P}\right)^{\prime}\nonumber \\
 & -3\vartheta^{\prime}\left(\mathcal{J}^{\prime\prime}-\frac{64\pi}{9k}\mathcal{J}\mathcal{P}\right)-5\vartheta^{\prime\prime}\mathcal{J}^{\prime}-\frac{10}{3}\vartheta^{\prime\prime\prime}\mathcal{J}-\frac{2}{3}\varrho\left(\mathcal{P}^{\prime\prime}-\frac{8\pi}{k}\mathcal{P}^{2}\right)^{\prime}\nonumber \\
 & -3\varrho^{\prime}\left(\mathcal{P}^{\prime\prime}-\frac{32\pi}{9k}\mathcal{P}^{2}\right)-5\mathcal{P}^{\prime}\varrho^{\prime\prime}-\frac{10}{3}\mathcal{P}\varrho^{\prime\prime\prime}+\frac{k}{6\pi}\varrho^{(5)}\,,\nonumber 
\end{align}
while the parameters of the asymptotic symmetries must satisfy the
following ``deformed chirality conditions\textquotedblright , given
by 
\begin{align}
\dot{Y} & =Y^{\prime}\mu-Y\mu^{\prime}+Z^{\prime\prime}\vartheta^{\prime}-Z^{\prime}\vartheta^{\prime\prime}-\frac{2}{3}\left(Z^{\prime\prime\prime}\vartheta-Z\vartheta^{\prime\prime\prime}\right)+\frac{32\pi}{3k}\mathcal{P}\left(Z^{\prime}\vartheta-Z\vartheta^{\prime}\right)\,,\nonumber \\
\dot{T} & =Y^{\prime}\xi-Y\xi^{\prime}+Z^{\prime\prime}\varrho^{\prime}-Z^{\prime}\varrho^{\prime\prime}+T^{\prime}\mu-T\mu^{\prime}+\vartheta^{\prime}X^{\prime\prime}-\vartheta^{\prime\prime}X^{\prime}\nonumber \\
 & +\frac{2}{3}\left(Z\varrho^{\prime\prime\prime}-Z^{\prime\prime\prime}\varrho+\vartheta^{\prime\prime\prime}X-\vartheta X^{\prime\prime\prime}\right)\nonumber \\
 & +\frac{32\pi}{3k}\left[\mathcal{J}\left(Z^{\prime}\vartheta-Z\vartheta^{\prime}\right)+\mathcal{P}\left(Z^{\prime}\varrho-Z\varrho^{\prime}+\vartheta X^{\prime}-\vartheta^{\prime}X\right)\right]\,,\label{chira0}\\
\dot{Z} & =2Y^{\prime}\vartheta-Y\vartheta^{\prime}+Z^{\prime}\mu-2Z\mu^{\prime}\,,\nonumber \\
\dot{X} & =2Y^{\prime}\varrho-Y\varrho^{\prime}+Z^{\prime}\xi-2Z\xi^{\prime}-2\mu^{\prime}X+\mu X^{\prime}-T\vartheta^{\prime}+2T^{\prime}\vartheta\ ,\nonumber 
\end{align}
where dot corresponds to the derivative along $u$.

\bigskip{}

In sum, the extended asymptotic behaviour is described by gauge fields
of the form (\ref{A-Flat}), with
\begin{equation}
a_{(0)}=a_{\varphi}d\varphi+a_{u}du\ ,\label{a(0)-Flat}
\end{equation}
where $a_{\varphi}$ and $a_{u}$ are given by eqs. (\ref{a-phi-flat})
and (\ref{au}), respectively.

As explained in \cite{HPTT}, \cite{BHPTT}, the canonical generators
depend only on $a_{\varphi}$, and never on the chemical potentials,
so that their expression is precisely the same as the one obtained
in \cite{GMPT}, i.e., 
\begin{equation}
Q_{(0)}\left(T,Y,Z,X\right)=-\int\left(T\mathcal{P}+Y\mathcal{J}+Z\mathcal{V}+X\mathcal{W}\right)d\varphi\ .\label{Q_flat}
\end{equation}
Hence, by construction, the asymptotic symmetries are still generated
by the higher spin extension of the centrally-extended BMS$_{3}$
algebra, whose mode expansion is explicitly written in eq. (\ref{HS_BMS})
below.

As an ending remark of this section, according to (\ref{feqs_flat}),
it is fairly clear that configurations for which the fields $\mathcal{P}$,
$\mathcal{J}$, $\mathcal{V}$, $\mathcal{W}$, as well as their corresponding
chemical potentials $\xi$, $\mu$, $\vartheta$, $\varrho$, are
constant, solve the field equations. This class of solutions then
explicitly reads
\begin{equation}
a_{(0)}=\left[J_{1}+\frac{2\pi}{k}(\mathcal{J}P_{0}+\mathcal{P}J_{0})+\frac{\pi}{k}(\mathcal{V}P_{00}+\mathcal{W}J_{00})\ \right]d\varphi+a_{u}\left(\xi,\mu,\vartheta,\varrho\right)du\,,\label{a_flat_cosmology}
\end{equation}
with
\begin{align}
a_{u}\left(\xi,\mu,\vartheta,\varrho\right) & =\frac{2\pi}{k}\left(\mu\mathcal{J}+\xi\mathcal{P}+2\vartheta\mathcal{V}+2\varrho\mathcal{W}\right)P_{0}+\xi P_{1}+\frac{2\pi}{k}\left(\mu\mathcal{P}+2\vartheta\mathcal{W}\right)J_{0}+\mu J_{1}\nonumber \\
 & +\frac{\pi}{k}\left(\mu\mathcal{V}+\xi\mathcal{W}+\frac{8}{k}\vartheta\mathcal{J}\mathcal{P}+\frac{4\pi}{k}\varrho\mathcal{P}^{2}\right)P_{00}+\frac{4\pi}{k}\left(\varrho\mathcal{P}+\vartheta\mathcal{J}\right)P_{01}+\varrho P_{11}\label{au_flat_cosmol}\\
 & +\frac{\pi}{k}\left(\mu\mathcal{W}+\frac{4\pi}{k}\vartheta\mathcal{P}^{2}\right)J_{00}+\frac{4\pi}{k}\vartheta\mathcal{P}J_{01}+\vartheta J_{11}\,,\nonumber 
\end{align}
and provides the searched for higher spin extension of the locally
flat cosmological spacetimes \cite{Ezawa:1992nk},\cite{Cornalba:2002fi},\cite{Cornalba:2003kd},\cite{Barnich:2012xq},\cite{Bagchi:2012xr}.
As it follows from (\ref{Q_flat}), the solutions not only carry mass
and angular momentum, determined by the spin-$2$ charges $\mathcal{P}$,
$\mathcal{J}\,$, respectively, but they are also endowed with global
charges of spin $3$, determined through $\mathcal{W}$, $\mathcal{V}$.
These higher spin charges are of electric and magnetic type, i.e.,
they are even and odd under parity, respectively.

The analysis of different classes of cosmological spacetimes endowed
with higher spin fields has also been discussed in \cite{PTT-Chethan},
\cite{PTT-PZ}, \cite{Krishnan:2013tza-1}.

\section{Recovering the asymptotically flat structure from a vanishing cosmological
constant limit}

\label{section_limit}

In this section we explain how the extended asymptotically flat structure
described above can be recovered from a suitable vanishing cosmological
constant limit of the asymptotic conditions recently constructed in
\cite{HPTT}, \cite{BHPTT}, for the case that includes black holes
endowed with global higher spin charges. By means of a suitable modification
of the Lagrange multipliers, the latter set enlarges the one independently
proposed in \cite{Theisen-HS}, \cite{Henneaux-HS}, so as to accommodate
chemical potentials in a way that it is consistent with the extended
conformal symmetry, spanned by two copies of the W$_{3}$ algebra.

\subsection{Extended asymptotic behaviour and higher spin black holes with negative
cosmological constant}

\label{sec_negative_lambda}

In the case of negative cosmological constant, $\Lambda=-\frac{1}{\ell^{2}}$,
the theory is described by two independent $sl\left(3,\mathbb{R}\right)$
gauge fields, $A^{+}$ and $A^{-}$, so that the action reads
\begin{equation}
I=I_{CS}\left[A^{+}\right]-I_{CS}\left[A^{-}\right]\ .
\end{equation}
The level is now given by $\kappa=k\ell$, and the bracket corresponds
to a quarter of the trace in the fundamental representation of $sl\left(3,\mathbb{R}\right)$,
generated by $L_{i}$ and $W_{m}$, with $i=-1,0,1$, and $m=-2,-1,...,+2$.

As explained in \cite{HPTT}, \cite{BHPTT}, the radial dependence
can be consistently gauged away, i.e., $A^{\pm}=g^{-1}a^{\pm}g+g^{-1}dg$,
with $g=g(r)$, so that the asymptotic form of the gauge fields is
described through 
\begin{equation}
a^{\pm}=\left(L_{\pm1}^{\pm}-\frac{2\pi}{\kappa}\mathcal{L}^{\pm}L_{\mp1}^{\pm}-\frac{\pi}{2\kappa}\mathcal{W}^{\pm}W_{\mp2}^{\pm}\right)d\varphi+\lambda^{\pm}\left(\xi_{\pm},\eta_{\pm}\right)dt\ ,\label{a+-with  Lambda}
\end{equation}
where $\lambda^{\pm}\left(\xi_{\pm},\eta_{\pm}\right)$ is given by
\begin{align}
\lambda^{\pm}\left(\xi_{\pm},\eta_{\pm}\right) & =\pm\left[\xi_{\pm}L_{\pm1}^{\pm}+\eta_{\pm}W_{\pm2}^{\pm}\mp\xi_{\pm}^{\prime}L_{0}^{\pm}\mp\eta_{\pm}^{\prime}W_{\pm1}^{\pm}+\frac{1}{2}\left(\xi_{\pm}^{\prime\prime}-\frac{4\pi}{\kappa}\xi_{\pm}\mathcal{L}^{\pm}+\frac{8\pi}{\kappa}\mathcal{W}^{\pm}\eta_{\pm}\right)L_{\mp1}^{\pm}\right.\nonumber \\
 & -\left(\frac{\pi}{2\kappa}\mathcal{W}^{\pm}\xi_{\pm}+\frac{7\pi}{6k}\mathcal{L}^{\pm\prime}\eta_{\pm}^{\prime}+\frac{\pi}{3\kappa}\eta_{\pm}\mathcal{L}^{\pm\prime\prime}+\frac{4\pi}{3\kappa}\mathcal{L}^{\pm}\eta_{\pm}^{\prime\prime}\right.\left.-\frac{4\pi^{2}}{\kappa^{2}}(\mathcal{L}^{\pm})^{2}\eta_{\pm}-\frac{1}{24}\eta_{\pm}^{\prime\prime\prime\prime}\right)W_{\mp2}^{\pm}\nonumber \\
 & +\left.\frac{1}{2}\left(\eta_{\pm}^{\prime\prime}-\frac{8\pi}{\kappa}\mathcal{L}^{\pm}\eta_{\pm}\right)W_{0}^{\pm}\mp\frac{1}{6}\left(\eta_{\pm}^{\prime\prime\prime}-\frac{8\pi}{\kappa}\eta_{\pm}\mathcal{L}^{\pm\prime}-\frac{20\pi}{\kappa}\mathcal{L}^{\pm}\eta_{\pm}^{\prime}\right)W_{\mp1}^{\pm}\right]\ .
\end{align}
Here $\mathcal{L}^{\pm}$, $\mathcal{W}^{\pm}$, $\xi_{\pm}$, $\eta_{\pm}$
stand for arbitrary functions of time and the angular coordinate $t$,
$\varphi$. Note that the ``chemical potentials\textquotedblright \ $\xi_{\pm}$,
$\eta_{\pm}$, appear only in the components of the gauge fields along
time, and they are assumed to be fixed at infinity, i.e. $\delta\xi_{\pm}=\delta\eta_{\pm}=0$.

The asymptotic behaviour of the dynamical fields $a_{\varphi}^{\pm}$
is preserved under gauge transformations generated by $\lambda^{\pm}(\varepsilon_{\pm},\chi_{\pm})$,
where the parameters $\varepsilon_{\pm}$, $\chi_{\pm}$ are independent
functions of $t$, $\varphi$, provided the fields $\mathcal{L}^{\pm}$,
$\mathcal{W}^{\pm}$ transform as
\begin{align}
\delta\mathcal{L}^{\pm} & =\pm2\mathcal{L}^{\pm}\varepsilon_{\mathcal{\pm}}^{\prime}\pm\varepsilon_{\pm}\mathcal{L}^{\pm\prime}\mp\frac{\kappa}{4\pi}\varepsilon_{\mathcal{\pm}}^{\prime\prime\prime}\mp2\chi_{\pm}\mathcal{W}^{\pm\prime}\mp3\mathcal{W}^{\pm}\chi_{\pm}^{\prime}\ ,\label{eq:deltaL}\\
\delta\mathcal{W}^{\pm} & =\pm3\mathcal{W}^{\pm}\varepsilon_{\mathcal{\pm}}^{\prime}\pm\varepsilon_{\pm}\mathcal{W}^{\pm\prime}\pm\frac{2}{3}\chi_{\pm}\left(\mathcal{L}^{\pm\prime\prime\prime}-\frac{16\pi}{\kappa}\left(\mathcal{L}^{\pm}\right)^{2\prime}\right)\pm3\left(\mathcal{L}^{\pm\prime\prime}-\frac{64\pi}{9\kappa}\left(\mathcal{L}^{\pm}\right)^{2}\right)\chi_{\pm}^{\prime}\nonumber \\
 & \pm5\chi_{\pm}^{\prime\prime}\mathcal{L}^{\pm\prime}\pm\frac{10}{3}\mathcal{L}^{\pm}\chi_{\pm}^{\prime\prime\prime}\mp\frac{\kappa}{12\pi}\chi_{\pm}^{\left(5\right)}\ .\label{eq:deltaW}
\end{align}
It is then simple to verify that the fall-off of the Lagrange multipliers
$a_{t}^{\pm}$ is maintained under the asymptotic symmetries, provided
the field equations are fulfilled at the asymptotic region, i.e.,
\begin{align}
\dot{\mathcal{L}}^{\pm} & =\pm2\mathcal{L}^{\pm}\xi_{\mathcal{\pm}}^{\prime}\pm\xi_{\pm}\mathcal{L}^{\pm\prime}\mp\frac{\kappa}{4\pi}\xi_{\mathcal{\pm}}^{\prime\prime\prime}\mp2\eta_{\pm}\mathcal{W}^{\pm\prime}\mp3\mathcal{W}^{\pm}\eta_{\pm}^{\prime}\ ,\label{eq:LPuntoW3}\\
\dot{\mathcal{W}}^{\pm} & =\pm3\mathcal{W}^{\pm}\xi_{\mathcal{\pm}}^{\prime}\pm\xi_{\pm}\mathcal{W}^{\pm\prime}\pm\frac{2}{3}\eta_{\pm}\left(\mathcal{L}^{\pm\prime\prime\prime}-\frac{16\pi}{\kappa}\left(\mathcal{L}^{\pm}\right)^{2\prime}\right)\pm3\left(\mathcal{L}^{\pm\prime\prime}-\frac{64\pi}{9\kappa}\left(\mathcal{L}^{\pm}\right)^{2}\right)\eta_{\pm}^{\prime}\nonumber \\
 & \pm5\eta_{\pm}^{\prime\prime}\mathcal{L}^{\pm\prime}\pm\frac{10}{3}\mathcal{L}^{\pm}\eta_{\pm}^{\prime\prime\prime}\mp\frac{\kappa}{12\pi}\eta_{\pm}^{\left(5\right)}\ ,\label{eq:WPuntoW3}
\end{align}
while the parameters have to satisfy the following ``deformed chirality
conditions\textquotedblright :
\begin{align}
\dot{\varepsilon_{\pm}} & =\pm\left(\varepsilon_{\pm}\xi_{\pm}^{\prime}-\xi_{\pm}\varepsilon_{\pm}^{\prime}\right)\pm\left(\chi_{\pm}^{\prime}\eta_{\pm}^{\prime\prime}-\chi_{\pm}^{\prime\prime}\eta_{\pm}^{\prime}\right)\pm\frac{2}{3}\left(\chi_{\pm}^{\prime\prime\prime}\eta_{\pm}-\chi_{\pm}\eta_{\pm}^{\prime\prime\prime}\right)\pm\frac{2^{6}\pi}{3\kappa}\left(\chi_{\pm}\eta_{\pm}^{\prime}-\eta_{\pm}\chi_{\pm}^{\prime}\right)\mathcal{L}^{\pm}\,,\nonumber \\
\dot{\chi_{\pm}} & =\mp\left(\chi_{\pm}^{\prime}\xi_{\pm}-2\chi_{\pm}\xi_{\pm}^{\prime}\right)\pm\left(\varepsilon_{\pm}\eta_{\pm}^{\prime}-2\varepsilon_{\pm}^{\prime}\eta_{\pm}\right)\,.\label{eq:chiPunto}
\end{align}
Hence, by construction, the canonical generators associated to the
asymptotic symmetries do not depend on the chemical potentials, and
are given by 
\begin{equation}
Q_{\pm}\left(\varepsilon_{\pm},\chi_{\pm}\right)=\pm\int\left(\varepsilon_{\pm}\mathcal{L}_{\pm}-\chi_{\pm}\mathcal{W}_{\pm}\right)d\varphi\ ,\label{PTT-Q-W3}
\end{equation}
so that their Poisson brackets span two copies of the W$_{3}$ algebra
with the standard central extension, $c=\frac{3\ell}{2G}$ \cite{Brown-Henneaux}.

As explained in \cite{HPTT}, \cite{BHPTT}, configurations with constant
fields $\mathcal{L}^{\pm}$, $\mathcal{W}^{\pm}$, and chemical potentials
$\xi_{\pm}$, $\eta_{\pm}$, given by
\begin{equation}
a^{\pm}=\left(L_{\pm1}^{\pm}-\frac{2\pi}{\kappa}\mathcal{L}^{\pm}L_{\mp1}^{\pm}-\frac{\pi}{2\kappa}\mathcal{W}^{\pm}W_{\mp2}^{\pm}\right)d\varphi+\lambda^{\pm}\left(\xi_{\pm},\eta_{\pm}\right)dt\ ,\label{eq:W3-Asympt}
\end{equation}
with
\begin{align}
\lambda^{\pm}\left(\xi_{\pm},\eta_{\pm}\right) & =\pm\left[\xi_{\pm}L_{\pm1}^{\pm}+\eta_{\pm}W_{\pm2}^{\pm}-\frac{2\pi}{\kappa}\left(\xi_{\pm}\mathcal{L}^{\pm}-2\mathcal{W}^{\pm}\eta_{\pm}\right)L_{\mp1}^{\pm}\right.\nonumber \\
 & \left.-\left(\frac{\pi}{2\kappa}\mathcal{W}^{\pm}\xi_{\pm}-\frac{4\pi^{2}}{\kappa^{2}}(\mathcal{L}^{\pm})^{2}\eta_{\pm}\right)W_{\mp2}^{\pm}-\frac{4\pi}{\kappa}\left(\mathcal{L}^{\pm}\eta_{\pm}\right)W_{0}^{\pm}\right]\ ,\label{LambdaMN-W3}
\end{align}
manifestly solve the field equations (\ref{eq:LPuntoW3}), (\ref{eq:WPuntoW3}),
and describe black holes solutions carrying not only mass and angular
momentum, but also nontrivial spin-$3$ charges of electric and magnetic
type.

\subsection{A suitable gauge choice to obtain the extended asymptotic behaviour
in the vanishing cosmological constant limit}

As explained in \cite{Limite-plano}, even in the case of pure gravity,
the limiting process that allows to recover the asymptotically flat
behaviour of the metric from the Brown-Henneaux boundary conditions,
in a way that it is consistent with the asymptotic BMS$_{3}$ symmetry
\cite{Barnich:2006av},\cite{Barnich:2010eb}, turns out to be a very
subtle one.\ Hence, in order to show how the whole extended asymptotic
structure described in section \ref{section-flat conditions} can
be obtained from the one in \ref{sec_negative_lambda} in the vanishing
cosmological constant limit, we follow a similar strategy as the one
implemented in \cite{GMPT} and \cite{BDMT}, for the cases of higher
spin gravity (without chemical potentials) and supergravity, respectively.

The procedure consists in finding a suitable gauge choice that allows
to take the limit in a straightforward way. As explained in \cite{BHPTT},
the searched for gauge choice must be ``permissible\textquotedblright ,
in the sense that it should not interfere with the asymptotic symmetry
algebra. Although not strictly necessary, one of the simplest possibilities
is looking for a gauge choice that does not depend on the global charges,
since this ensures that the allowed gauge transformations commute
with the variation of the canonical generators.

Let us then consider the extended asymptotic conditions for the theory
with negative cosmological constant, described by the connections
$a^{\pm}$ in eq. (\ref{a+-with Lambda}). Our goal can then be achieved
expressing the fall-off of the entire gauge field with the following
permissible gauge choice:
\begin{equation}
a_{(\Lambda)}:=a^{+}+g^{-1}a^{-}g\ ,\label{a-Lambda-new gauge}
\end{equation}
where $g$ stands for a constant group element, given by 
\begin{equation}
g=e^{\frac{\pi}{2}\left(L_{1}^{-}+L_{-1}^{-}\right)}\,.\label{g cachilupi}
\end{equation}
It is worth pointing out that this gauge choice is even simpler than
the ones performed in \cite{GMPT} and \cite{BDMT}, since it only
affects one of the copies of the connection. Indeed, the effect of
this gauge transformation on $a^{-}$ just amounts to modify its components
according to $L_{i}\rightarrow(-1)^{i+1}L_{-i}$, and $W_{m}\rightarrow(-1)^{m}W_{-m}$.

It is then useful to perform the following change of basis
\[
J_{0}^{\pm}=-\frac{1}{2}L_{-1}^{\pm},\ \ \ J_{2}^{\pm}=L_{0}^{\pm},\ \ \ J_{1}^{\pm}=L_{1}^{\pm},
\]
\begin{equation}
T_{00}^{\pm}=-\frac{1}{4}W_{-2}^{\pm}\ ,\ \ \ T_{02}^{\pm}=\frac{1}{2}W_{-1}^{\pm}\ ,\ \ \ T_{22}^{\pm}=-W_{0}^{\pm}\ ,\ \ \ T_{12}^{\pm}=-W_{1}^{\pm}\ ,\ \ \ T_{11}^{\pm}=-W_{2}^{\pm}\ ,
\end{equation}
being equivalent to the one in \cite{GMPT}, up to an automorphism,
so that the generators $T_{ab}$ become traceless. This change of
basis is then followed by
\begin{equation}
J_{a}^{\pm}=\frac{J_{a}\pm\ell P_{a}}{2}\ \ \ ;\ \ \ T_{ab}^{\pm}=\frac{J_{ab}\pm\ell P_{ab}}{2}\ ,
\end{equation}
so that the $sl(3,R)\oplus sl(3,R)$ generators are now described
by the set $\left\{ J_{a},P_{a},J_{ab},P_{ab}\right\} $, and the
algebra reads \cite{Theisen-HS} 
\begin{align}
\lbrack J_{a},J_{b}] & =\epsilon_{abc}J^{c}\ \ ;\ \ [J_{a},P_{b}]=\epsilon_{abc}P^{c}\ \ ;\ \ [P_{a},P_{b}]=-\Lambda\epsilon_{abc}J^{c}\ ,\label{HS_Lambda-Poincaroid}\\
\lbrack J_{a},J_{bc}] & =\epsilon_{\;\; a(b}^{m}J_{c)m}\ \ ;\ \ [J_{a},P_{bc}]=\epsilon_{\;\; a(b}^{m}P_{c)m}\ \ ;\ \ [P_{a},J_{bc}]=\epsilon_{\;\; a(b}^{m}P_{c)m}\ \ ;\ \ [P_{a},P_{bc}]=-\Lambda\epsilon_{\;\; a(b}^{m}J_{c)m}\ ,\nonumber \\
\lbrack J_{ab},J_{cd}] & =-\eta_{\left(a\right\vert (c}\epsilon_{d)\left\vert b\right)m}J^{m}\ \ ;\ \ [J_{ab},P_{cd}]=-\eta_{\left(a\right\vert (c}\epsilon_{d)\left\vert b\right)m}P^{m}\ \ ;\ \ [P_{ab},P_{cd}]=\Lambda\eta_{\left(a\right\vert (c}\epsilon_{d)\left\vert b\right)m}J^{m}\ .\nonumber 
\end{align}
It is also natural and convenient to redefine the fields and chemical
potentials according to
\begin{equation}
\mathcal{L}^{\pm}=\frac{\ell\mathcal{P\pm J}}{2}\,\,\,;\,\,\,\mathcal{W}^{\pm}=\frac{\ell\mathcal{W\pm V}}{2}\ ,\label{redef-fields}
\end{equation}
and
\begin{equation}
\xi_{\pm}=\frac{\xi}{\ell}\pm\mu\,\,\,;\,\,\,\eta_{\pm}=-\left(\frac{\varrho}{\ell}\pm\vartheta\right)\ ,\label{redef-chem}
\end{equation}
respectively; as well as renaming the time coordinate as $t=u$. The
asymptotic form of the gauge field (\ref{a-Lambda-new gauge}) then
reduces to
\begin{equation}
a_{(\Lambda)}=a_{(0)}-\frac{2\pi\Lambda}{k}\Xi\ du\ ,\label{a(Lambda)-Chi}
\end{equation}
where $a_{(0)}$ acquires the same form as in eq. (\ref{a(0)-Flat}),
and
\begin{align}
\Xi & :=\left(\xi\mathcal{J}+2\varrho\mathcal{V}\right)J_{0}+2\varrho\mathcal{J}J_{01}-\frac{2}{3}\left(\varrho\mathcal{J}^{\prime}+\frac{5}{2}\varrho^{\prime}\mathcal{J}\right)J_{02}\label{Chi}\\
 & +\frac{1}{2}\left(\xi\mathcal{V}+\frac{4\pi}{k}\vartheta\mathcal{J}^{2}+\frac{8\pi}{k}\varrho\mathcal{J}\mathcal{P}-\frac{7}{3}\varrho^{\prime}\mathcal{J}^{\prime}-\frac{2}{3}\varrho\mathcal{J}^{\prime\prime}-\frac{8}{3}\varrho^{\prime\prime}\mathcal{J}\right)J_{00}+\frac{2\pi}{k}\varrho\mathcal{J}^{2}P_{00}\,.\nonumber 
\end{align}
At this step, it must be emphasized that the generators in eq. (\ref{a(Lambda)-Chi})
fulfill the algebra (\ref{HS_Lambda-Poincaroid}), with $\Lambda\neq0$.
Consistency then requires that the gauge parameters have to be redefined
accordingly with the chemical potentials, i.e.,
\begin{equation}
\epsilon_{\pm}=\frac{T}{\ell}\pm Y\,\,\,;\,\,\,\chi_{\pm}=-\left(\frac{X}{\ell}\pm Z\right)\ ,\label{redef-params}
\end{equation}
so that the connection (\ref{a(Lambda)-Chi}) is preserved by gauge
transformations spanned by
\begin{equation}
\eta_{(\Lambda)}=\eta_{(0)}-\frac{2\pi\Lambda}{k}\Xi(T,Y,X,Z)\ du\ ,\label{eta(Lambda)}
\end{equation}
where $\eta_{(0)}$ has the same form as in eq. (\ref{lambda tilda}),
provided the fields transform as
\begin{align}
\delta_{(\Lambda)}\mathcal{J} & =\delta_{(0)}\mathcal{J}\ ,\nonumber \\
\delta_{(\Lambda)}\mathcal{P} & =\delta_{(0)}\mathcal{P}-\Lambda\left[T\mathcal{J}^{\prime}+2\mathcal{J}T^{\prime}+2X\mathcal{V}^{\prime}+3\mathcal{V}X^{\prime}\right]\ ,\nonumber \\
\delta_{(\Lambda)}\mathcal{W} & =\delta_{(0)}\mathcal{W}-\Lambda\left[T\mathcal{V}^{\prime}+3T^{\prime}\mathcal{V}-3X^{\prime}\mathcal{J}^{\prime\prime}-5X^{\prime\prime}\mathcal{J}^{\prime}-\frac{2}{3}X\mathcal{J}^{\prime\prime\prime}-\frac{10}{3}\mathcal{J}X^{\prime\prime\prime}\right.\label{deltas(Lambda)-fields}\\
 & \left.+\frac{16\pi}{3k}\left(Z\left(\mathcal{J}^{2}\right)^{\prime}+2Z^{\prime}\mathcal{J}^{2}+2X\left(\mathcal{P}\mathcal{J}\right)^{\prime}+4X^{\prime}\mathcal{P}\mathcal{J}\right)\right]\ ,\nonumber \\
\delta_{(\Lambda)}\mathcal{V} & =\delta_{(0)}\mathcal{V}-\frac{16\pi}{3k}\Lambda\left[X\left(\mathcal{J}^{2}\right)^{\prime}+2X^{\prime}\mathcal{J}^{2}\right]\ ,\nonumber 
\end{align}
where $\delta_{(0)}\mathcal{J}$, $\delta_{(0)}\mathcal{P}$, $\delta_{(0)}\mathcal{W}$,
$\delta_{(0)}\mathcal{V}$ are given by eq. (\ref{delta0}).

The field equations (\ref{eq:LPuntoW3}), (\ref{eq:WPuntoW3}), as
well as the deformed chirality conditions in (\ref{eq:chiPunto})
now read
\begin{align}
\dot{\mathcal{J}} & =2\mu^{\prime}\mathcal{J}+\mu\mathcal{J}^{\prime}+\xi\mathcal{P}^{\prime}+2\xi^{\prime}\mathcal{P}-\frac{k}{2\pi}\xi^{\prime\prime\prime}+2\vartheta\mathcal{V}^{\prime}+3\vartheta^{\prime}\mathcal{V}+2\mathcal{W}^{\prime}\varrho+3\mathcal{W}\varrho^{\prime}\ ,\nonumber \\
\dot{\mathcal{P}} & =2\mu^{\prime}\mathcal{P}+\mu\mathcal{P}^{\prime}-\frac{k}{2\pi}\epsilon^{\prime\prime\prime}+2\vartheta\mathcal{W}^{\prime}+3\vartheta^{\prime}\mathcal{W}-\Lambda\left[\xi\mathcal{J}^{\prime}+2\mathcal{J}\xi^{\prime}+2\varrho\mathcal{V}^{\prime}+3\mathcal{V}\varrho^{\prime}\right]\ ,\nonumber \\
\mathcal{\dot{W}} & =3\mu^{\prime}\mathcal{W}+\mu\mathcal{W}^{\prime}-\frac{2}{3}\vartheta\left(\mathcal{P}^{\prime\prime}-\frac{8\pi}{k}\mathcal{P}^{2}\right)^{\prime}-3\vartheta^{\prime}\left(\mathcal{P}^{\prime\prime}-\frac{32\pi}{9k}\mathcal{P}^{2}\right)-5\vartheta^{\prime\prime}\mathcal{P}^{\prime}-\frac{10}{3}\vartheta^{\prime\prime\prime}\mathcal{P}+\frac{k}{6\pi}\vartheta^{\left(5\right)}\nonumber \\
 & -\Lambda\left[\xi\mathcal{V}^{\prime}+3\xi^{\prime}\mathcal{V}-3\varrho^{\prime}\mathcal{J}^{\prime\prime}-5\varrho^{\prime\prime}\mathcal{J}^{\prime}-\frac{2}{3}\varrho\mathcal{J}^{\prime\prime\prime}-\frac{10}{3}\mathcal{J}\varrho^{\prime\prime\prime}\right.\label{Feqs-(Lambda)}\\
 & \left.+\frac{16\pi}{3k}\left(\vartheta\left(\mathcal{J}^{2}\right)^{\prime}+2\vartheta^{\prime}\mathcal{J}^{2}+2\varrho\left(\mathcal{P}\mathcal{J}\right)^{\prime}+4\varrho^{\prime}\mathcal{P}\mathcal{J}\right)\right]\ ,\nonumber \\
\dot{\mathcal{V}} & =3\mu^{\prime}\mathcal{V}+\mu\mathcal{V}^{\prime}+\xi\mathcal{W}^{\prime}+3\xi^{\prime}\mathcal{W}-\frac{2}{3}\vartheta\left(\mathcal{J}^{\prime\prime}-\frac{16\pi}{k}\mathcal{J}\mathcal{P}\right)^{\prime}-3\vartheta^{\prime}\left(\mathcal{J}^{\prime\prime}-\frac{64\pi}{9k}\mathcal{J}\mathcal{P}\right)-5\vartheta^{\prime\prime}\mathcal{J}^{\prime}\nonumber \\
 & -\frac{10}{3}\vartheta^{\prime\prime\prime}\mathcal{J}-\frac{2}{3}\varrho\left(\mathcal{P}^{\prime\prime}-\frac{8\pi}{k}\mathcal{P}^{2}\right)^{\prime}-3\varrho^{\prime}\left(\mathcal{P}^{\prime\prime}-\frac{32\pi}{9k}\mathcal{P}^{2}\right)-5\mathcal{P}^{\prime}\varrho^{\prime\prime}-\frac{10}{3}\mathcal{P}\varrho^{\prime\prime\prime}+\frac{k}{6\pi}\varrho^{(5)}\nonumber \\
 & -\frac{16\pi}{3k}\Lambda\left[\varrho\left(\mathcal{J}^{2}\right)^{\prime}+2\varrho^{\prime}\mathcal{J}^{2}\right]\ ,\nonumber 
\end{align}
and
\begin{align}
\dot{Y} & =Y^{\prime}\mu-Y\mu^{\prime}+Z^{\prime\prime}\vartheta^{\prime}-Z^{\prime}\vartheta^{\prime\prime}-\frac{2}{3}\left(Z^{\prime\prime\prime}\vartheta-Z\vartheta^{\prime\prime\prime}\right)+\frac{32\pi}{3k}\mathcal{P}\left(Z^{\prime}\vartheta-Z\vartheta^{\prime}\right)-\Lambda\left[T^{\prime}\xi-T\xi^{\prime}\right.\nonumber \\
 & \left.-\varrho^{\prime\prime}X^{\prime}+\varrho^{\prime}X^{\prime\prime}+\frac{2}{3}\left(\varrho^{\prime\prime\prime}X-\varrho X^{\prime\prime\prime}\right)+\frac{32\pi}{3k}\left(\mathcal{J}\left(Z^{\prime}\varrho-Z\varrho^{\prime}-\vartheta^{\prime}X+\vartheta X^{\prime}\right)+\mathcal{P}\left(\varrho X^{\prime}-\varrho^{\prime}X\right)\right)\right]\,,\nonumber \\
\dot{T} & =Y^{\prime}\xi-Y\xi^{\prime}+Z^{\prime\prime}\varrho^{\prime}-Z^{\prime}\varrho^{\prime\prime}+T^{\prime}\mu-T\mu^{\prime}+\vartheta^{\prime}X^{\prime\prime}-\vartheta^{\prime\prime}X^{\prime}+\frac{2}{3}\left(Z\varrho^{\prime\prime\prime}-Z^{\prime\prime\prime}\varrho+\vartheta^{\prime\prime\prime}X-\vartheta X^{\prime\prime\prime}\right)\nonumber \\
 & +\frac{32\pi}{3k}\left(\mathcal{J}\left(Z^{\prime}\vartheta-Z\vartheta^{\prime}\right)+\mathcal{P}\left(Z^{\prime}\varrho-Z\varrho^{\prime}+\vartheta X^{\prime}-\vartheta^{\prime}X\right)\right)-\frac{32\pi}{3k}\Lambda\left[\mathcal{J}\left(\varrho X^{\prime}-\varrho^{\prime}X\right)\right]\,,\label{chira(Lambda)}\\
\dot{Z} & =2Y^{\prime}\vartheta-Y\vartheta^{\prime}+Z^{\prime}\mu-2Z\mu^{\prime}+\Lambda\left[T\varrho^{\prime}-2T^{\prime}\varrho-\xi X^{\prime}+2\xi^{\prime}X\right]\ ,\nonumber \\
\dot{X} & =2Y^{\prime}\varrho-Y\varrho^{\prime}+Z^{\prime}\xi-2Z\xi^{\prime}-2\mu^{\prime}X+\mu X^{\prime}-T\vartheta^{\prime}+2T^{\prime}\vartheta\,,\nonumber 
\end{align}
respectively.

Note that since the gauge choice in (\ref{a-Lambda-new gauge}), (\ref{g cachilupi})
is permissible, the global charges do not change, i.e., 
\[
Q_{(\Lambda)}\left(T,Y,Z,X\right)=Q_{+}\left(\varepsilon_{+},\chi_{+}\right)-Q_{-}\left(\varepsilon_{-},\chi_{-}\right)\ ,
\]
which by virtue of the redefinitions in eqs. (\ref{redef-fields}),
(\ref{redef-params}), they now read
\begin{equation}
Q_{(\Lambda)}\left(T,Y,Z,X\right)=Q_{(0)}\left(T,Y,Z,X\right)\ ,\label{Q(Lambda)}
\end{equation}
whose expression remarkably agrees with the one for $\Lambda=0$ in
eq. (\ref{Q_flat}). Nonetheless, it must be emphasized that here
the fields transform according to (\ref{deltas(Lambda)-fields}),
instead of (\ref{delta0}). Therefore, once expanded in Fourier modes
according to $X=\frac{1}{2\pi}\sum_{m}X_{m}e^{im\theta}$, the algebra
of the canonical generators, given by two copies of W$_{3}$, now
reads
\begin{align}
i\{\mathcal{J}_{n},\mathcal{J}_{m}\} & =(n-m)\mathcal{J}_{n+m}\ \ \ ;\ \ \ i\{\mathcal{P}_{n},\mathcal{P}_{m}\}=-\Lambda(n-m)\mathcal{J}_{n+m}\ ,\nonumber \\
i\{\mathcal{J}_{n},\mathcal{P}_{m}\} & =(n-m)\mathcal{P}_{n+m}+kn^{3}\delta_{m+n}\ ,\nonumber \\
i\{\mathcal{P}_{n},\mathcal{W}_{m}\} & =-\Lambda(2n-m)\mathcal{V}_{n+m}\ \ \ ;\ \ \ i\{\mathcal{P}_{n},\mathcal{V}_{m}\}=(2n-m)\mathcal{W}_{n+m}\ ,\nonumber \\
i\{\mathcal{J}_{n},\mathcal{W}_{m}\} & =(2n-m)\mathcal{W}_{n+m}\ \ \ ;\ \ \ i\{\mathcal{J}_{n},\mathcal{V}_{m}\}=(2n-m)\mathcal{V}_{n+m}\ ,\label{algebra(Lambda)}\\
i\{\mathcal{W}_{n},\mathcal{W}_{m}\} & =-\frac{\Lambda}{3}(n-m)(2n^{2}+2m^{2}-mn)\mathcal{J}_{m+n}-\frac{16\Lambda}{3k}(n-m)\sum_{j}\mathcal{J}_{j}\mathcal{P}_{n+m-j}\ ,\nonumber \\
i\{\mathcal{W}_{n},\mathcal{V}_{m}\} & =\frac{1}{3}(n-m)(2n^{2}+2m^{2}-mn)\mathcal{P}_{m+n}+\frac{8}{3k}(n-m)\Omega_{m+n}+\frac{k}{3}n^{5}\delta_{m+n}\ ,\nonumber \\
i\{\mathcal{V}_{n},\mathcal{V}_{m}\} & =\frac{1}{3}(n-m)(2n^{2}+2m^{2}-mn)\mathcal{J}_{m+n}+\frac{16}{3k}(n-m)\sum_{j}\mathcal{P}_{j}\mathcal{J}_{n+m-j}\ ,\nonumber 
\end{align}
where
\begin{equation}
\Omega_{n}=\sum_{m}\left(\mathcal{P}_{n-m}\mathcal{P}_{m}-\Lambda\mathcal{J}_{n-m}\mathcal{J}_{m}\right)\ .
\end{equation}

\subsubsection{Taking the $\Lambda\rightarrow0$ limit}

The limiting process that allows to recover the whole structure of
the vanishing cosmological constant case, can then be taken in a very
transparent way. Indeed, when $\Lambda=-\frac{1}{\ell^{2}}\rightarrow0$,
the generators of the $sl(3,%TCIMACRO{\U{211d}}%
%BeginExpansion
\mathbb{R}%EndExpansion
)\oplus sl(3,%TCIMACRO{\U{211d}}%
%BeginExpansion
\mathbb{R}%EndExpansion
)$ algebra (\ref{HS_Lambda-Poincaroid}) clearly span their corresponding
Inönü-Wigner contraction (\ref{HS_Poincare}). Hence, according to
(\ref{a(Lambda)-Chi}), the gauge fields with the special gauge choice
in (\ref{a-Lambda-new gauge}) fulfill $a_{(\Lambda)}\rightarrow a_{(0)}$,
i.e., they manifestly reduce to their flat counterpart in eq. (\ref{a(0)-Flat}).
Analogously, eq. (\ref{eta(Lambda)}) implies that $\eta_{(\Lambda)}\rightarrow\eta_{(0)}$,
so that the gauge transformations that preserve the asymptotic form
of $a_{(0)}$ in (\ref{lambda tilda}), are also recovered in the
limit. This is also the case of the transformation law of the fields,
since when $\Lambda\rightarrow0$, eq. (\ref{delta0}) is readily
obtained from (\ref{deltas(Lambda)-fields}). Moreover, according
to eqs. (\ref{Feqs-(Lambda)}) and (\ref{chira(Lambda)}), the field
equations and the chirality conditions in the flat case, given by
(\ref{feqs_flat}), (\ref{chira0}), respectively, are also recovered.

Noteworthy, as expressed by (\ref{Q(Lambda)}), since $Q_{(\Lambda)}=Q_{(0)}$,
the expression for the canonical generators is automatically obtained
without the need of taking the limit.

Finally, by virtue of (\ref{algebra(Lambda)}), it is also clear that
the higher spin extension of the conformal symmetry, spanned by two
copies of the W$_{3}$ algebra reduces to the higher spin extension
of the BMS$_{3}$ algebra, given by
\begin{align}
i\{\mathcal{J}_{n},\mathcal{J}_{m}\} & =(n-m)\mathcal{J}_{n+m}\ \ \ ;\ \ \ i\{\mathcal{P}_{n},\mathcal{P}_{m}\}=0\ ,\nonumber \\
i\{\mathcal{J}_{n},\mathcal{P}_{m}\} & =(n-m)\mathcal{P}_{n+m}+kn^{3}\delta_{m+n}\ ,\nonumber \\
i\{\mathcal{P}_{n},\mathcal{W}_{m}\} & =0\ \ \ ;\ \ \ i\{\mathcal{P}_{n},\mathcal{V}_{m}\}=(2n-m)\mathcal{W}_{n+m}\ ,\nonumber \\
i\{\mathcal{J}_{n},\mathcal{W}_{m}\} & =(2n-m)\mathcal{W}_{n+m}\ \ \ ;\ \ \ i\{\mathcal{J}_{n},\mathcal{V}_{m}\}=(2n-m)\mathcal{V}_{n+m}\ ,\label{HS_BMS}\\
i\{\mathcal{W}_{n},\mathcal{W}_{m}\} & =0\,\nonumber \\
i\{\mathcal{W}_{n},\mathcal{V}_{m}\} & =\frac{1}{3}(n-m)(2n^{2}+2m^{2}-mn)\mathcal{P}_{m+n}+\frac{8}{3k}(n-m)\Omega_{m+n}+\frac{k}{3}n^{5}\delta_{m+n}\ ,\nonumber \\
i\{\mathcal{V}_{n},\mathcal{V}_{m}\} & =\frac{1}{3}(n-m)(2n^{2}+2m^{2}-mn)\mathcal{J}_{m+n}+\frac{16}{3k}(n-m)\sum_{j}\mathcal{P}_{j}\mathcal{J}_{n+m-j}\ ,\nonumber 
\end{align}
where
\begin{equation}
\Omega_{n}=\sum_{m}\mathcal{P}_{n-m}\mathcal{P}_{m}\ .
\end{equation}

As an ending remark of this section, it is apparent that, since it
has been shown that the whole asymptotic structure fulfills $a_{(\Lambda)}\rightarrow a_{(0)}$
in the vanishing $\Lambda$ limit, the higher spin black hole solution
(\ref{eq:W3-Asympt}) reduces to the higher spin extension of locally
flat cosmological spacetimes in eq. (\ref{a_flat_cosmology}).

\section{Higher spin extension of locally flat cosmological spacetimes and
thermodynamics}

\label{HScosmo+thermo}

As explained in section \ref{section-flat conditions}, the generalized
asymptotic conditions (\ref{a(0)-Flat}) naturally accommodate a higher
spin extension of cosmological spacetimes, given by (\ref{a_flat_cosmology}).
The solution is explicitly described not only in terms of their global
spin-$2$ and spin-$3$ charges, but also by their corresponding chemical
potentials, which are strictly necessary in order to have a regular
Euclidean configuration. In this case, from the metric formalism,
it can be inferred that the topology of the three-dimensional manifold
turns out to be the one of a solid torus, but with a reversed orientation
as compared with the case of black holes \cite{Carlip:1994gc}. This
is explained in the appendix (see fig. 1). Note also that, as explained
in \cite{HPTT}, \cite{BHPTT}, since all the chemical potentials
are manifestly incorporated along the temporal components of the gauge
fields, the analysis can be carried out for a fixed range of the angular
coordinates of the torus, i.e., we assume that $0<\tau\leq1$, and
$0<\varphi\leq2\pi$. This is particularly useful in the case of higher
spin gravity, since the torus clearly has not enough room to accommodate
all the chemical potentials in the range of coordinates.

The entropy can then be obtained from the following expression
\begin{align}
S & =\frac{k}{2\pi}\left[\int_{r_{+}}d\tau d\varphi\left\langle A_{\tau}A_{\varphi}\right\rangle \right]_{\text{on-shell}}\nonumber \\
 & =k\left[\left\langle a_{\tau}a_{\varphi}\right\rangle \right]_{\text{on-shell}}\ .\label{Entropy-generic}
\end{align}
In the microcanonical ensemble, this is the boundary term that is
needed so that the action acquires a bona fide extremum. The field
equations then have to hold everywhere, which implies that the fields
have to be regular at the horizon. This procedure then ensures that
the first law is also fulfilled either in the canonical or in the
grand canonical ensembles.

It is worth highlighting that neither the Poincaré algebra nor their
higher spin extensions, as in eq. (\ref{HS_Lambda-Poincaroid}), admit
a suitable standard matrix representation from which the Casimir operators,
and so invariant bilinear form (\ref{CS3}) that is required to construct
the action, can be recovered from the trace of a product of the generators.
Consequently, regularity of the Euclidean solution, which is guaranteed
by requiring the holonomy along the thermal circle to be trivial,
cannot be straightforwardly implemented through its diagonalization.

In the next subsection, we describe a general procedure that allows
to implement the regularity condition without the need of an explicit
matrix representation of the entire gauge group, but only of its Lorentz-like
subgroup.

In few words, we show that the temporal components of generalized
dreibeins $e_{\tau}$, can be consistently gauged away, which partially
fixes the chemical potentials; so that the remaining conditions can
be obtained by requiring the holonomy of the generalized spin connection
$\omega$ along a thermal circle to be trivial.

The procedure aforementioned then allows to carry out the analysis
of the thermodynamic properties in a direct way. As a warming up exercise,
this is first performed in the case of pure gravity, and then we show
how the analysis extends once the higher spin charges and their corresponding
chemical potentials are switched on.

\subsection{Procedure to implement the regularity conditions in a generic form}

When one deals with locally flat connections defined on a solid torus,
since the thermal circles $\mathcal{C}$ are contractible, regularity
of the fields $a=g^{-1}dg$, implies the triviality of their holonomies
along them, i.e.,
\begin{equation}
H_{\mathcal{C}}=P\exp\left[%TCIMACRO{\dint_{\mathcal{C}}}%
%BeginExpansion
{\displaystyle \int_{\mathcal{C}}}%EndExpansion
a_{\mu}dx^{\mu}\right]=\exp\left[%TCIMACRO{\dint_{0}^{1}}%
%BeginExpansion
{\displaystyle \int_{0}^{1}}%EndExpansion
a_{\tau}d\tau\right]=g^{-1}(\tau)g(\tau+1)=I_{c}\ ,\label{Hc}
\end{equation}
where $I_{c}$ stands for a suitable element of the center of the
group. If the gauge group admits an appropriate matrix representation,
the regularity conditions can then be directly implemented through
the diagonalization of $H_{\mathcal{C}}$. Alternatively, according
to (\ref{Hc}), one could obtain the explicit form of the gauge group
element $g$, so that regularity implies that $g$ is well defined
along along the cycle. Hence, in a suitable patch around the origin,
the ``regularizing gauge transformation\textquotedblright , generated
by $g^{-1}$, makes the temporal component of the gauge fields to
vanish, i.e., $a_{\tau}=0$.

Note that in pure gravity, as well as in higher spin gravity, the
gauge group always possesses the following structure (see, e.g., \cite{Theisen-HS})
\begin{equation}
\lbrack J,J]\sim J\ \ ;\ \ [J,P]\sim P\ \ ;\ \ [P,P]\sim-\Lambda\ J\ ,\label{generic gauge group}
\end{equation}
where $J$ stand for the Lorentz-like generators, and according to
the value of $\Lambda$, the generators $P$ correspond to the extended
translations, or (A)dS boosts. The connection is then generically
of the form
\[
a=\omega J+eP\ ,
\]
being locally flat, i.e., $a=g^{-1}dg$, where the group element $g$,
by virtue of (\ref{generic gauge group}), can always be written as
\[
g=g_{P}\cdot g_{J}\ ,
\]
with $g_{P}:=e^{\lambda P}$, and $g_{J}:=e^{\Theta J}$.

Therefore, the regularity condition of the fields can always be implemented
in a ``hybrid way\textquotedblright \ as follows:

\bigskip{}

$\mathbf{(i)}$ Finding the group element $g_{P}$ that allows to
gauge away the temporal components of the gauge field along $P$,
so that one can consistently set $e_{\tau}=0$. This partially fixes
the chemical potentials.

\bigskip{}

$\mathbf{(ii)}$ The remaining conditions can then be implemented
through the diagonalization of the holonomy matrix associated to the
spin connection along the thermal circle;\ i.e., without the need
of finding the explicit form of $g_{J}$.

\bigskip{}

Note that in the case of a finite number of fields with spin $s=1$,
$2$, ..., $N$, since the Lorentz-like group is given by $SL(N,%TCIMACRO{\U{211d}}%
%BeginExpansion
\mathbb{R}%EndExpansion
)$, the holonomy associated to $\omega_{\tau}$ always admits a reducible
matrix representation that allows to diagonalize it.

It is worth pointing out that if the radial dependence were brought
back, it is clear that the regularity conditions had to be imposed
at the horizon. The procedure explained above certainly can always
be applied in any case. Indeed, it is simple to verify that the regularity
conditions for the gauge fields that describe black holes in the case
of $\Lambda<0$, which fix the chemical potentials in terms of the
global charges, are successfully reproduced in this way. It should
then be emphasized that this procedure becomes particularly useful
when the gauge group does not admit a suitable matrix representation,
so that the Casimir operators, and hence the invariant bilinear form
(\ref{CS3}), cannot be recovered from the trace of a product of the
generators, as it is the case of $\Lambda=0$.

\subsection{Warming up with pure gravity}

\label{warming up pure gravity}

For the sake of simplicity, let us first consider the case of pure
gravity with vanishing cosmological constant, so that the gauge group
is spanned by the Poincaré algebra, whose Lorentz subalgebra is given
by $sl(2,%TCIMACRO{\U{211d}}%
%BeginExpansion
\mathbb{R}%EndExpansion
)$. Hence, the field configuration that describes cosmological spacetimes
(\ref{a_flat_cosmology}) can be obtained from (\ref{a_flat_cosmology}),
in the case of vanishing higher spin charges and their corresponding
chemical potentials; i.e., for $\mathcal{V\,}=\mathcal{W}=\vartheta=\varrho=0$.
The connection then reads
\begin{align}
a_{(0)} & =\left(J_{1}+\frac{2\pi}{k}\mathcal{J}P_{0}+\frac{2\pi}{k}\mathcal{P}J_{0}\right)d\varphi\label{a(0).pure grav}\\
 & +\left[\mu\left(J_{1}+\frac{2\pi}{k}\mathcal{P}J_{0}\right)+\xi P_{1}+\frac{2\pi}{k}\left(\mu\mathcal{J}+\xi\mathcal{P}\right)P_{0}\right]du\,,\nonumber 
\end{align}
and hence, according to (\ref{Entropy-generic}) the entropy is readily
found to be given by
\begin{equation}
S=4\pi\left[\xi\mathcal{P}+\mu\mathcal{J}\right]_{\text{on-shell}}\ ,\label{S-Smarr-grav}
\end{equation}
where the chemical potentials have to fulfill the regularity conditions.
According to the procedure described above, the first step $(i)$
consists in finding a suitable gauge transformation $g_{P}$ that
allows to consistently gauge away the temporal components of the dreibein,
i.e., $e_{u}=0$. It is simple to see that the required permissible
group element is of the form
\begin{equation}
g_{P}=e^{\lambda_{2}P_{2}}\,,\label{gp-grav}
\end{equation}
so that the gauge field now reads
\begin{align*}
a & =\left(J_{1}+\lambda_{2}P_{1}+\frac{2\pi}{k}\left(\mathcal{J}-\lambda_{2}\mathcal{P}\right)P_{0}+\frac{2\pi}{k}\mathcal{P}J_{0}\right)d\varphi\\
 & +\left[\mu\left(J_{1}+\frac{2\pi}{k}\mathcal{P}J_{0}\right)+\left(\mu\lambda_{2}+\xi\right)P_{1}+\frac{2\pi}{k}\left(\mu\mathcal{J}-\left(\mu\lambda_{2}-\xi\right)\mathcal{P}\right)P_{0}\right]du\,.
\end{align*}
Hence, the dreibein component $e_{u}^{1}$ vanishes if
\[
\lambda_{2}=-\frac{\xi}{\mu}\,,
\]
while the remaining component $e_{u}^{0}$ also does provided the
following condition is fulfilled:
\[
\mu=-2\xi\frac{\mathcal{P}}{\mathcal{J}}\,.
\]
In this gauge, the connection is then explicitly given by
\[
a=\left(J_{1}-\frac{\xi}{\mu}P_{1}+\frac{2\pi}{k}\left(\mathcal{J}+\frac{\xi}{\mu}\mathcal{P}\right)P_{0}+\frac{2\pi}{k}\mathcal{P}J_{0}\right)d\varphi-2\xi\frac{\mathcal{P}}{\mathcal{J}}\left(J_{1}+\frac{2\pi}{k}\mathcal{P}J_{0}\right)du\,.
\]
It is worth to remark that the regularizing group element $g_{P}$
in (\ref{gp-grav}) is non singular and globally well-defined.

The remaining step $(ii)$ amounts to require the holonomy of the
spin connection along the thermal circle to be trivial. In the fundamental
representation of $sl(2,%TCIMACRO{\U{211d}}%
%BeginExpansion
\mathbb{R}%EndExpansion
)$, this condition reduces to
\[
tr\left[\left(\omega_{\tau}\right)^{2}\right]+2\pi^{2}=2\pi^{2}-\frac{8\pi}{k}\frac{\xi^{2}\mathcal{P}^{3}}{\mathcal{J}^{2}}=0\,,
\]
being solved by
\begin{equation}
\xi^{2}=\frac{\pi k\mathcal{J}^{2}}{4\mathcal{P}^{3}}\,.\label{Chi squared}
\end{equation}
Note that since we are dealing with a cosmological horizon, the orientation
of the solid torus is reversed as compared with the one of the black
hole, so that the chemical potential $\xi$ corresponds to the minus
branch of (\ref{Chi squared}). This goes by hand with the positivity
of the Hawking temperature, since $\xi=-\frac{1}{T}$.

In sum, the regularity conditions imply that the chemical potentials
become fixed in terms of the global charges according to
\[
\mu=sgn(\mathcal{J})\sqrt{\frac{\pi k}{\mathcal{P}}}\ \ \ ,\ \ \ \xi=-\frac{\sqrt{\pi k}\left\vert \mathcal{J}\right\vert }{2\mathcal{P}^{3/2}}\,,
\]
which allows to express the entropy in terms of the global charges
as
\begin{equation}
S=2\pi\sqrt{\frac{\pi k}{\mathcal{P}}}\left\vert \mathcal{J}\right\vert \,.\label{S-pure grav}
\end{equation}
This result agrees with the one for General Relativity, i.e., $S=\frac{A}{4G}$
(see appendix), which in the metric formalism, was explicitly carried
out in \cite{Barnich:2012xq}, \cite{Bagchi:2012xr}.

\subsection{Switching on higher spin charges and chemical potentials}

\label{switching on HS charges and chem pots}

Here we show that the thermodynamic analysis of the higher spin extension
of the cosmological spacetimes (\ref{a_flat_cosmology}) proceeds
as explained above in a straightforward way. Indeed, in this case
the entropy (\ref{Entropy-generic}) evaluates as
\begin{equation}
S=2\pi\left[2\xi\mathcal{P}+2\mu\mathcal{J}+3\varrho\mathcal{W}+3\vartheta\mathcal{V}\right]_{\text{on-shell}}\ ,\label{Soffshell}
\end{equation}
where the chemical potentials have to satisfy the regularity conditions.
In order to solve them, let us apply the first step $(i)$, which
consists in finding the gauge transformation $g_{P}$ that allows
to gauge away the components of the dreibein along time. The permissible
globally well-defined gauge transformation is found to be given by
\begin{equation}
g_{P}=e^{\hat{\lambda}}\,,\label{gp-grav}
\end{equation}
with
\begin{equation}
\hat{\lambda}=\frac{(32\pi\mathcal{J}\mathcal{P}^{2}-9k\mathcal{V}\mathcal{W})P_{2}+10\pi\mathcal{P}(2\mathcal{P}\mathcal{V}-3\mathcal{J}\mathcal{W})P_{02}+(6k\mathcal{P}\mathcal{V}-9k\mathcal{J}\mathcal{W})P_{12}}{64\pi\mathcal{P}^{3}-27k\mathcal{W}^{2}}\ ,
\end{equation}
so that the temporal components of the dreibein vanish provided
\begin{align}
\xi & =\frac{32\pi\mathcal{P}(3\mathcal{J}\mathcal{W}-2\mathcal{P}\mathcal{V})\vartheta+\left(9k\mathcal{V}\mathcal{W}-32\pi\mathcal{J}\mathcal{P}^{2}\right)\mu}{64\pi\mathcal{P}^{3}-27k\mathcal{W}^{2}}\ ,\label{hol_xi}\\
\varrho & =\frac{\left(18k\mathcal{V}\mathcal{W}-64\pi\mathcal{J}\mathcal{P}^{2}\right)\vartheta+\left(9k\mathcal{J}\mathcal{W}-6k\mathcal{P}\mathcal{V}\right)\mu}{64\pi\mathcal{P}^{3}-27k\mathcal{W}^{2}}\ .\label{hol_rho}
\end{align}
The entropy (\ref{Soffshell}) then simplifies as
\begin{equation}
S=2\pi\left[\mu\mathcal{J}+\vartheta\mathcal{V}\right]_{\text{on-shell}}\ .\label{Sfinal}
\end{equation}
Step $(ii)$ is then implemented through requiring the holonomy of
the spin connection along the thermal circle to be trivial. Since
the Lorentz-like group is now given by $sl(3,%TCIMACRO{\U{211d}}%
%BeginExpansion
\mathbb{R}%EndExpansion
)$, the remaining conditions reduce to
\begin{align}
tr\left[\left(\omega_{t}\right)^{2}\right]+8\pi^{2} & =\frac{8\pi}{k}\mathcal{P}\mu^{2}+\frac{24\pi}{k}\mathcal{W}\vartheta\mu+\frac{128\pi^{2}}{3k^{2}}\mathcal{P}^{2}\vartheta^{2}-8\pi^{2}=0\ ,\label{at2}\\
tr\left[\left(\omega_{t}\right)^{3}\right] & =\frac{\pi}{k}\mathcal{W}\mu^{3}+\frac{32\pi^{2}}{3k^{2}}\mathcal{P}^{2}\vartheta\mu^{2}+\frac{16\pi^{2}}{k^{2}}\mathcal{P}\mathcal{W}\vartheta^{2}\mu+\frac{16\pi^{2}}{k^{2}}\mathcal{W}^{2}\vartheta^{3}-\frac{512\pi^{3}}{27k^{3}}\mathcal{P}^{3}\vartheta^{3}=0\ ,\label{at3}
\end{align}
which do not depend on $\mathcal{J}$, $\mathcal{V}$.

Conditions (\ref{at2}), (\ref{at3}) admit different branches of
solutions. In order to make contact with the cosmological configurations
in the case of pure gravity, it is convenient to consider the following
classes of solutions
\begin{align}
\mu & =\pm\sqrt{\frac{\pi k}{\mathcal{P}}}\cos\left(\frac{2\Phi}{3}\right)\sec\left(\Phi\right)\ ,\label{hol_mu}\\
\vartheta & =\frac{\sqrt{3}k}{4\mathcal{P}}\sin\left(\frac{\Phi}{3}\right)\sec\left(\Phi\right)\ ,\label{hol_theta}
\end{align}
with
\begin{equation}
\sin(\Phi)=\mp\frac{3}{8}\sqrt{\frac{3k}{\pi\mathcal{P}^{3}}}\mathcal{W}\ ,\label{PHI grande}
\end{equation}
so that consistency implies that the sign in (\ref{hol_mu}) coincides
with the one of the angular momentum $\mathcal{J}$. Hence, the entropy
can be expressed in terms of the global charges according to
\begin{equation}
S=2\pi\sqrt{\frac{\pi k}{\mathcal{P}}}\sec\left(\Phi\right)\left[\left\vert \mathcal{J}\right\vert \cos\left(\frac{2\Phi}{3}\right)+\sqrt{\frac{3k}{\pi\mathcal{P}}}\frac{\mathcal{V}}{4}\sin\left(\frac{\Phi}{3}\right)\right]\ .\label{Entropy}
\end{equation}
Note that the branch that is continuously connected with the cosmological
spacetime of General Relativity ($\Phi=0$) corresponds to $-\frac{\pi}{2}<\Phi<\frac{\pi}{2}$.
The advantage of writing the entropy in terms of the ``angular variable\textquotedblright \ $\Phi$,
is that it also holds for different branches.

As a final remark, in addition to $\mathcal{P}>0$, the bound that
guarantees that the entropy is a real function, directly comes from
(\ref{PHI grande}), which is given by
\[
\left\vert \mathcal{W}\right\vert \leq\frac{8}{3\sqrt{3}}\sqrt{\frac{\pi}{k}}\mathcal{P}^{\frac{3}{2}}\ .
\]
When this bound saturates, the configuration is ``extremal\textquotedblright \ in
the sense that the holonomy of the generalized spin connection along
the thermal circle is no longer trivial, because there is a change
in the topology ($%TCIMACRO{\U{211d}}%
%BeginExpansion
\mathbb{R}%EndExpansion
\times S^{1}\times S^{1}$). Note that in the branch that is connected with the pure gravity
solution, positivity of the entropy implies that the angular momentum
has to be bounded from below according to
\[
\left\vert \mathcal{J}\right\vert >-\sqrt{\frac{3k}{\pi\mathcal{P}}}\frac{\mathcal{V}\sin\left(\frac{\Phi}{3}\right)}{4\cos\left(\frac{2\Phi}{3}\right)}\ ,
\]
so that the bound becomes nontrivial provided the sign of the spin-$3$
charge of electric type $\mathcal{W}$ is the opposite of its magnetic-like
counterpart $\mathcal{V}$. This is the analogue of the condition
that guarantees the existence of an event horizon in the case of pure
gravity (see eq. \ref{horizon_pure_gravity}).

\section{Final remarks}

\label{final remarks}

The extension of the generalized asymptotically flat behaviour to
the case of spins $s\geq2$ can be directly performed along the lines
of \cite{HPTT}, \cite{BHPTT}. Thus, in order to incorporate the
chemical potentials, one begins with the asymptotic behaviour at a
fixed time slice $u=u_{0}$. The radial dependence can also be gauged
away by a group element of the form $h(r)=e^{-rP_{0}}$, and hence,
as described in \cite{GMPT}, the asymptotic form of the spacelike
connection is given by
\begin{equation}
a_{\varphi}=J_{1}+\frac{2\pi}{k}(\mathcal{J}P_{0}+\mathcal{P}J_{0})+\frac{\pi}{k}(\mathcal{V}_{3}P_{00}+\mathcal{W}_{3}J_{00})+\frac{\pi}{k}(\mathcal{V}_{4}P_{000}+\mathcal{W}_{4}J_{000})+...\label{a-phi-N}
\end{equation}
where the spin-$s$ generators $P_{a_{1}\cdot\cdot\cdot a_{s-1}}$,
$J_{a_{1}\cdot\cdot\cdot a_{s-1}}$, are assumed to be fully symmetric
and traceless, and the fields $\mathcal{M}$, $\mathcal{J}$, $\mathcal{W}_{3}$,
$\mathcal{V}_{3}$, $\mathcal{W}_{4}$, $\mathcal{V}_{4}$, ..., stand
for arbitrary functions of $u_{0}$, $\varphi$. It was also shown
in \cite{GMPT} that the asymptotic behaviour (\ref{a-phi-N}) can
be consistently recovered from the vanishing $\Lambda$ limit of its
AdS$_{3}$ analogue \cite{Henneaux-HS}, \cite{Theisen-HS}, \cite{PTT-CFP},
after a suitable permissible gauge choice.

The asymptotic form of the connection is then maintained under gauge
transformations of the form $\delta a=d\eta+\left[a,\eta\right]$,
where $\eta=\eta\left(T,Y,Z_{3},X_{3},Z_{4},X_{4},...\right)$ depends
on arbitrary parameters of $u_{0}$, $\varphi$, provided the fields
transform in a suitable way. Consequently, since the dynamical fields
evolve in time through a gauge transformation generated by $a_{u}$,
the asymptotic symmetries will be preserved along time evolution provided
the Lagrange multiplier belongs to the allowed family, i.e.,
\begin{equation}
a_{u}=\eta\left(\xi,\mu,\vartheta_{3},\varrho_{3},\vartheta_{4},\varrho_{4},...\right)\,,\label{au-N}
\end{equation}
where $\xi,\mu,\vartheta_{3},\varrho_{3},\vartheta_{4},\varrho_{4},$
... stand for arbitrary functions of time and the angular coordinates
that are assumed to be fixed at the boundary, and describe the chemical
potentials conjugated to the corresponding global charges. Preserving
the asymptotic form of $a_{u}$ then implies that the field equations
hold at the asymptotic region, and also provides a precise set of
conditions for the parameters that generate the asymptotic symmetries.

The asymptotically flat behaviour in the case of spins $s\geq2$,
is then described by gauge fields of the form 
\begin{equation}
a=a_{\varphi}d\varphi+a_{u}du\ ,\label{a(0)-Flat}
\end{equation}
with $a_{\varphi}$ and $a_{u}$ given by eqs. (\ref{a-phi-N}) and
(\ref{au-N}), respectively.

By construction, since the surface integrals that describe the canonical
generators depend only on $a_{\varphi}$, and not on the chemical
potentials, their expression coincides with the one that can be obtained
from \cite{GMPT},
\begin{equation}
Q\left(T,Y,Z_{3},X_{3},Z_{4},X_{4},...\right)=-\int\left(T\mathcal{P}+Y\mathcal{J}+Z_{3}\mathcal{V}_{3}+X_{3}\mathcal{W}_{3}\mathcal{+}Z_{4}\mathcal{V}_{4}+X_{4}\mathcal{W}_{4}\mathcal{+}\cdot\cdot\cdot\right)d\varphi\ ,\label{Q_flat-N}
\end{equation}
and hence, the algebra of the asymptotic symmetries is still generated
by the corresponding higher spin extension of the centrally-extended
BMS$_{3}$ algebra.

The locally flat cosmologies can then be readily extended to the case
that includes chemical potentials and higher spin charges of spin
$s\geq2$. Indeed, this is the case when the fields $\mathcal{M}$,
$\mathcal{J}$, $\mathcal{W}_{3}$, $\mathcal{V}_{3}$, $\mathcal{W}_{4}$,
$\mathcal{V}_{4}$, ..., and the arbitrary functions $\xi,\mu,\vartheta_{3},\varrho_{3},\vartheta_{4},\varrho_{4},$
... are constant.

\bigskip{}

It is also worth pointing out that a different possible extension
of our results could be carried out along a different front. For instance,
note that the flat analogue of the well-known result that allows to
describe AdS$_{3}$ gravity with Brown-Henneaux boundary conditions
in terms of a Liouville theory at the boundary \cite{Cousaert-Henneaux-V},
has recently been constructed in \cite{Liouville-dual-plano}; and
it has also been shown that the dual theory at null infinity can be
consistently recovered from the vanishing $\Lambda$ limit of its
AdS$_{3}$ counterpart \cite{Liouville-bms}. Thus, following these
lines, and assuming that the asymptotically flat behaviour of gravity
coupled to spin-$3$ fields is described as in \cite{ABFGR}, \cite{GMPT},
the higher spin extension of the dual theory at null infinity was
recently shown to correspond to a flat analogue of Toda theory \cite{Gonzalez:2014tba}.
It would then be interesting to explore how the dual theory becomes
modified once the generalized asymptotically flat behaviour, described
in section \ref{section-flat conditions} is taken into account, as
well as carrying out the extension to fields of spin $s\geq2$.

\bigskip{}

As a final remark, we would like to make a comparison with the results
that have been recently reported by Gary, Grumiller, Riegler and Rosseel
in \cite{Gary:2014ppa}. Since the generalized asymptotically flat
behaviour in the case of spin-$3$ gravity also follows the lines
of \cite{HPTT}, \cite{BHPTT}, we naturally agree. Nonetheless, in
order to make the link with the vanishing $\Lambda$ limit, the path
they follow makes use of a prescription introduced by Krishnan, Raju
and Roy in \cite{PTT-Grassmann-Flat}. The prescription requires the
use of certain $6\times6$ matrix representation constructed out of
Grassmann variables, along with the introduction of different notions
of ``twisted\textquotedblright \ and ``hatted\textquotedblright \ traces
that allow to recover the metric and the spin-$3$ field, as well
as the invariant bilinear tensor that defines the bracket, respectively.
Besides, the regularity conditions of the fields are also obtained
in a completely different approach. Indeed, in order to carry out
the computation, they use another ($9\times9$) matrix representation,
followed by a prescription that partially relies on the vanishing
$\Lambda$ limit. Therefore, taking into account that we have followed
radically different approaches, it is very reassuring to check that
our results for the entropy of the higher spin extension of the cosmological
spacetimes agree in the cases that were considered in \cite{Gary:2014ppa}.
This can be explicitly seen as follows: if one restricts our entropy
formula (\ref{Entropy}) to the branch that is connected to the pure
gravity case, once the global charges are mapped according to 
\[
\mathcal{P}=\frac{k}{4\pi}\mathcal{\hat{M}\ \ \ };\ \ \ \mathcal{J}=\frac{k}{2\pi}\mathcal{\hat{L}\ \ \ };\ \ \ \mathcal{W}=\frac{2k}{\pi}\mathcal{\hat{V}}\ \ \ ;\ \ \ \mathcal{V}=\frac{4k}{\pi}\mathcal{\hat{U}\ },
\]
so that the variables $\mathcal{\hat{R}}$ and $\mathcal{\hat{P}}$
become
\[
\frac{\mathcal{\hat{R}}-1}{\mathcal{\hat{R}}^{3/2}}=\frac{1}{4}\sqrt{\frac{k}{\pi}}\frac{\mathcal{W}}{\mathcal{P}^{3/2}}\ \ \ ;\ \ \ \mathcal{\hat{P}}=\frac{1}{16}\sqrt{\frac{k}{\pi}}\frac{\mathcal{V}}{\mathcal{J}\sqrt{\mathcal{P}}}\ ,
\]
it reduces to the corresponding expression given in \cite{Gary:2014ppa}.
Here we have used a hat in order to distinguish their variables from
ours.

\bigskip{}

\acknowledgments We thank G. Barnich, L. Donnay, H. González, M.
Henneaux, M. Pino, and C. Troessaert, and especially to O. Fuentealba
for useful comments and enlightening discussions. J.M. and R.T. also
wish to thank the organizers of the ``Meeting on the horizon\textquotedblright ,
hosted by Pontificia Universidad Católica de Valparaíso, during March
2014, for the opportunity of presenting this work. This research has
been partially supported by Fondecyt grants N${^{\circ}}$ 1130658,
1121031, 11130260, 11130262, 3150448. Centro de Estudios Científicos
(CECs) is funded by the Chilean Government through the Centers of
Excellence Base Financing Program of Conicyt.

\appendix
%dummy comment inserted by tex2lyx to ensure that this paragraph is not empty

\section{Contact with the cosmological spacetime metric}

In order to recover the cosmological spacetime metric in the case
of pure gravity with $\Lambda=0$, one has to restore the radial dependence
of the gauge fields. It is useful to consider the following gauge
choice: 
\[
g_{\rho}=e^{\rho P_{2}}\,,
\]
so that the full gauge field now reads
\[
A=\omega^{a}J_{a}+e^{a}P_{a}=g_{\rho}^{-1}a_{(0)}g_{\rho}+g_{\rho}^{-1}dg_{\rho}\ ,
\]
where $a_{(0)}$ is given by (\ref{a(0).pure grav}). The connection
then reduces to
\begin{align*}
A & =\left(J_{1}+\rho P_{1}+\frac{2\pi}{k}\left(\mathcal{J}-\rho\mathcal{P}\right)P_{0}+\frac{2\pi}{k}\mathcal{P}J_{0}\right)d\varphi\\
 & +\left[\mu\left(J_{1}+\frac{2\pi}{k}\mathcal{P}J_{0}\right)+\left(\mu\rho+\xi\right)P_{1}+\frac{2\pi}{k}\left(\mu\mathcal{J}-\left(\mu\rho-\xi\right)\mathcal{P}\right)P_{0}\right]dt+P_{2}d\rho\,,
\end{align*}
and hence, the spacetime metric
\[
ds^{2}=\eta_{ab}e_{\mu}^{a}e_{\nu}^{b}dx^{\mu}dx^{\nu}\ ,
\]
is directly obtained in Schwarzschild-like coordinates,
\begin{equation}
ds^{2}=-\frac{4\pi}{k}\left(\frac{\pi\mathcal{J}^{2}}{kr^{2}}-\mathcal{P}\right)\xi^{2}dt^{2}+\frac{k}{4\pi}\left(\frac{\pi\mathcal{J}^{2}}{kr^{2}}-\mathcal{P}\right)^{-1}dr^{2}+r^{2}\left[\left(\mu+\frac{2\pi\mathcal{J}\xi}{kr^{2}}\right)dt+\text{\ensuremath{d\varphi}}\right]^{2}\,,\label{Lorentzian-cosmo-metric}
\end{equation}
where
\[
\rho=\frac{\mathcal{J}+\sqrt{\mathcal{J}^{2}-\frac{k}{\pi}\mathcal{P}r^{2}}}{2\mathcal{P}}\ ,
\]
which possesses a cosmological horizon located at 
\begin{equation}
r=r_{c}=\left\vert \mathcal{J}\right\vert \sqrt{\frac{\pi}{k\mathcal{P}}}>0\,.\label{horizon_pure_gravity}
\end{equation}
The Euclidean continuation of the cosmological spacetime metric is
recovered through $t\rightarrow-i\tau$, followed by
\[
\mathcal{P=-P}_{E}\ \ ;\ \ \mathcal{J}=i\mathcal{J}_{E}\ \ ;\ \ \xi=\xi_{E}\ \ ;\ \ \mu=i\mu_{E}\ \ ,
\]
so that the Euclidean metric reads
\begin{equation}
ds^{2}=\frac{4\pi}{k}\left(\mathcal{P}_{E}-\frac{\pi\mathcal{J}_{E}^{2}}{kr^{2}}\right)\xi^{2}d\tau^{2}+\frac{k}{4\pi}\left(\mathcal{P}_{E}-\frac{\pi\mathcal{J}_{E}^{2}}{kr^{2}}\right)^{-1}dr^{2}+r^{2}\left[\left(\mu_{E}+\frac{2\pi\mathcal{J}_{E}\xi}{kr^{2}}\right)d\tau+\text{\ensuremath{d\varphi}}\right]^{2}\,,\label{Euc-metric}
\end{equation}
This class of spaces was first discussed in \cite{Ezawa:1992nk},\cite{Cornalba:2002fi},\cite{Cornalba:2003kd},
and its thermodynamic properties have been thoroughly analyzed in
\cite{Barnich:2012xq},\cite{Bagchi:2012xr}.

It is worth emphasizing that here we have included the chemical potentials
explicitly in the metric, so that the range of the coordinates is
assumed to be fixed according to $0<\tau\leq1$, and $0<\varphi\leq2\pi$.
It is then simple to verify that the Euclidean metric (\ref{Euc-metric})
is regular at the horizon provided
\begin{align}
\mu_{E} & =\sqrt{\frac{\pi k}{\mathcal{P}}}\,,\\
\xi_{E} & =-\frac{\sqrt{\pi k}\left\vert \mathcal{J}\right\vert }{2\mathcal{P}^{3/2}}\,,\label{CHI euc}
\end{align}
while for at $r\rightarrow\infty$, asymptotically approaches to a
conical defect.

The topology of the Euclidean manifold can then be directly inferred
from the metric (\ref{Euc-metric}), which as shown in fig. 1, corresponds
to the one of a solid torus ($%TCIMACRO{\U{211d}}%
%BeginExpansion
\mathbb{R}%EndExpansion
^{2}\times S^{1}$), but with reversed orientation as compared with the one of the black
hole. Note that the cosmological horizon $r_{c}$ is located at the
``south pole\textquotedblright \ of the $r-\tau$ surface, and hence
the relationship between the ``chemical potential\textquotedblright \ $\xi$
and the Hawking temperature is given by $\xi=-\frac{1}{T}$, which
explains the use of the minus branch in (\ref{CHI euc}).

It is simple to verify that the entropy $S=\frac{A}{4G}$ agrees with
the one exclusively in terms of the gauge fields in eq. (\ref{S-pure grav}).

\newpage

\begin{figure}[H]
\begin{center}
\includegraphics[scale=0.4]{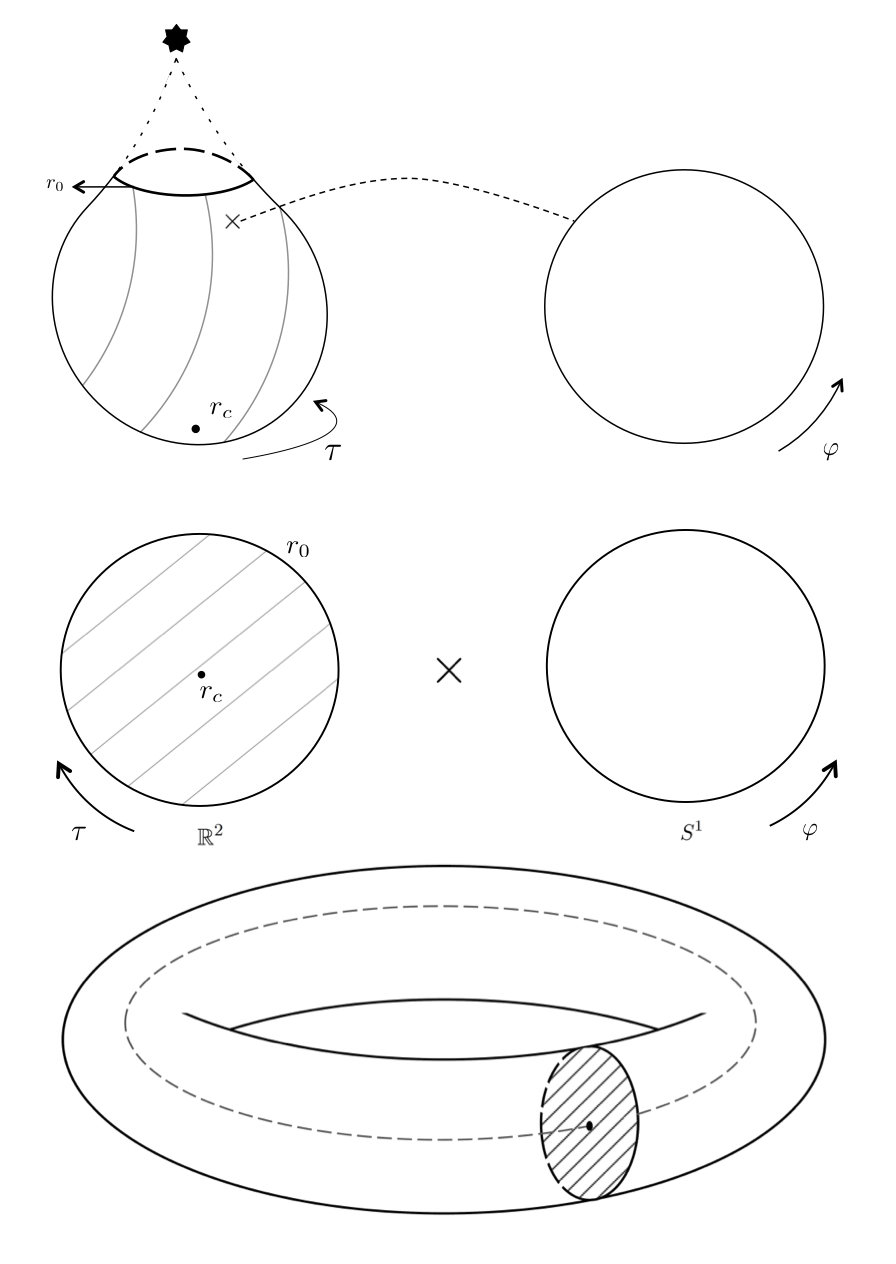}
\end{center}
\caption{The sequence shows that the topology of the Euclidean cosmological
spacetime coincides with the one of a black hole; i.e., it corresponds
to $%TCIMACRO{\U{211d}}%
%BeginExpansion
\mathbb{R}%EndExpansion
^{2}\times S^{1}$ (solid torus), but with reversed orientation (compare with fig. 1
of \cite{BHPTT}). The cosmological horizon $r_{c}$ is located at
the ``south pole\textquotedblright \ of the $r-\tau$ surface, which
asymptotically approaches to a conical defect at the tip of the drop,
so that a regulator at $r=r_{0}$ has to be introduced. Noncontractible
cycles then run along the circle $S^{1}$, being parametrized by the
angle $\varphi$.}
\end{figure}

\end{document}